\documentclass{aa}

\usepackage{graphicx}
\usepackage{txfonts}
\usepackage{mathrsfs} %
\usepackage[dvipsnames]{xcolor}

\makeatletter
\renewcommand*\aa@pageof{, page \thepage{} of \pageref*{LastPage}}
\makeatother

\usepackage[colorlinks=true, linkcolor=NavyBlue, citecolor=NavyBlue]{hyperref}
\usepackage[capitalise]{cleveref} %
\usepackage[separate-uncertainty=true, multi-part-units=single]{siunitx}

\graphicspath{{figures/}}
\DeclareGraphicsExtensions{.jpg, .pdf}

\newcommand{\code}[1]{\texttt{#1}\xspace}
\newcommand{\MAV}{\code{MPI-AMRVAC~2.2}\xspace}
\newcommand{\hd}{HD\xspace142527\xspace}
\newcommand{\Nabla}{\vec{\nabla}}
\newcommand{\Ms}{M_{\sun}}

\newcommand{\sub}{_\mathrm}

\newcommand{\vsym}{\langle v_\varphi \rangle}
\newcommand{\PA}{\mathrm{PA}}
\newcommand*\chem[1]{\ensuremath{\mathrm{#1}}}
\newlength{\figwidth}
\newlength{\figwidthsmall}
\begin{document}
\abstract
{Planets are formed amidst young circumstellar disks of gas and dust. The latter is traced
by thermal radiation, where strong asymmetric clumps were observed in a handful of cases.
These dust traps could be key to understand the early stages of planet formation, when
solids grow from micron-size to planetesimals.}
{Vortices are among the few known asymmetric dust trapping scenarios. The present work
aims at predicting their characteristics in a complementary observable. Namely,
line-of-sight velocities are well suited to trace the presence of a vortex. Moreover, the
dynamics of disks is subject to recent developments.}
{2D hydro-simulations were performed where a vortex forms at the edge of a gas depleted
region. We derived idealized line-of-sight velocity maps, varying disk temperature and
orientation relative to the observer.  The signal of interest, as a small perturbation to
the dominant axisymetric component in velocity, may be isolated in observational data
using a proxy for the dominant quasi-Keplerian velocity. We propose that the velocity
curve on the observational major axis be such a proxy.}
{Applying our method to the disk around \hd as a study case, we predict line-of-sight
velocities scarcely detectable by currently available facilities, depending on disk
temperature. We show that corresponding spirals patterns can also be detected with similar
spectral resolutions, which will help discriminating against alternative explanations.}
{} 

\title{Dynamical signatures of Rossby vortices in cavity-hosting disks}
\author{
  \href{https://orcid.org/0000-0001-8629-7068}{C.M.T. Robert}\inst{1}
  \and
  H. Méheut\inst{1}
  \and
  F. Ménard\inst{2}
}
\institute{
  Université Côte d'Azur, Observatoire de la Côte d'Azur, CNRS, Laboratoire Lagrange, Bd de l’Observatoire, CS 34229, 06304 Nice cedex 4,
  \textsc{France}
  \and
  Univ. Grenoble Alpes, CNRS, IPAG, F-38000 Grenoble,
  \textsc{France}
  \\
  \email{clement.robert@oca.eu}
}
\date{compiled \today; Received --; accepted --}
\titlerunning{All}

\maketitle
\section{Introduction}
\label{sec:intro}
Planets are formed in circumstellar disks made mainly of gas and some solid dust
components.  Many aspects of the processes implied in their formation remain challenging
to explain.  More specifically, the transition from small dust grains to large
planetesimals face two major obstacles: the drift barrier corresponding to fast inward
drifting due to gas headwind, and the collision barrier due to destructive collisions
\citep{Chiang2010a}.  Pressure bumps provide a solution to the drift barrier, as they act
as a barrier stopping the drifting solids and forming dust rings. Indeed, concentric dusty
rings are a common feature in resolved infrared images of protoplanetary disks
\citep{Andrews2018}.  Pressure bumps are also known to promote the formation of large
scale vortices, through to the Rossby Wave Instability (RWI), that are proposed as a
solution to the barriers in planetesimal formation. They both stop the dust drift and
harness efficient growth by lowering relative speeds between grains. This is why vortices
were proposed as a planet-promoting scenario \citep{Barge1995, Adams1995, Tanga1996,
Bracco1999}.  Moreover, it is well known that massive planets build up pressure bumps in
their vicinity, exciting vortex formation
\citep{deValBorro2006,deValBorro2007,Fu2014a,Hammer2017,Andrews2018,Baruteau2019}, which
in turn affects planetary migration \citep{Regaly2013,Ataiee2014,Mcnally2018}.  The study
of large vortices is thus key to understand planetary formation.

The RWI \citep{Lovelace1999, Li2000, Li2001} is a promising vortex-forming scenario, and
is expected where sharp density gradients are found. So-called "transitional" disks
provide such conditions at the outer edge of large ($\sim$ \SIrange{5}{100}{AU}) gas
cavities they host. Extensive computational effort has been dedicated to studying long
term evolution of RWI vortices \citep{Fu2014a, Meheut2012b, Regaly2017a, Andrews2018}.
Overall, eddies tend to form in a few tenth of orbital periods and survive for ${10^3}$ to
${10^4}$ orbital periods.

Concurrently, asymmetric dust crescents are being observed in thermal radiation of a
growing number of targets \citep{Cazzoletti2018, Dong2018, Isella2018, Casassus2019a,
Pineda2019} as well as in scattered emission \citep{Benisty2018}. Those clumps are
candidates for large vortices, and there have been attempts to explain their
formation as vortex-driven \citep{Regaly2012, Birnstiel2013}. Alternatively, disk
eccentricity \citep{Ataiee2013} and excitation by an eccentric companion \citep{Price2018}
were proposed to explain these azimuthal dust excess, however not reproducing the observed
dust-to-gas ratio.

Complementary measurements of the gas dynamics would be of great help in constraining and
rejecting concurrent explanations. Continuum emission traces the spatial
distribution of dust grains dynamically coupled with the gas, so it provides indirect
information on the underlying gas dynamics. However, direct measurements of the gas radial
velocity can now be achieved through Doppler-shifting of molecular lines, thanks to
increasingly sophisticated data reduction techniques \citep{Yen2016, Teague2016,
Teague2018}, and ever-enhanced spatial resolution \citep{Andrews2018}. It is
becoming possible to use these data to build connections to continuum asymmetries
\citep{Casassus2015b, Casassus2019} or searching for planet-induced deviations
\citep{Pinte2018, Teague2018a, Pinte2019a, Perez2020}.

Hence, observations in molecular line emission are key to confirm or reject current and
future vortex candidates.  The present paper is aimed at characterizing the dynamical
signatures expected for a single large Rossby eddy forming in the inner rim of a cavity,
by the means of hydro simulations.

The paper is organized as follows.  First, we describe the numerical setup of our hydro
simulations in \cref{sec:setup}.  We then provide insight on the observability of
resulting vortices and propose a method to extract their signature from observational data
in \cref{sec:signatures}.  Finally, we discuss the limits of our approach in
\cref{sec:discussion} and conclude in \cref{sec:ccl}.

\section{Hydro simulations setup}
\label{sec:setup}
Using \MAV \citep{Porth2014, Xia2018}, we perform 2D hydro simulations.
Namely, we solve Euler equations for an inviscid gas
\begin{align}
  \label{eq:mass_cons}
  &\partial_t \Sigma + \Nabla \cdot (\Sigma \vec{v}) = 0\:,
  \\
  \label{eq:navier-stokes-gas}
  &\left(\partial_t + \vec{v} \cdot  \Nabla \right) \Sigma \vec{v}
  = -\Sigma \Nabla \phi - \Nabla p \:,
\end{align}
where $\Sigma$ and $\vec{v}$ stand for surface density and velocity respectively,
$\phi\propto -1/r$
is a central gravitational potential and $p$ is the vertically integrated
pressure.
It is prescribed by a barotropic equation of state $p = S\Sigma^{\gamma}$ where
$S= 86.4$ code units\footnotemark characterizes the entropy and
$\gamma = 5/3$ is the adiabatic index. \footnotetext{Our code unit system is
completely described by mass, length and time normalisation constants respectively $m_* =
1$ solar mass, $r_* = \SI{100}{AU}$ and $t_* = 1$ orbital period of a test particle at
$r=r_j$\:.}
Sound speed is given as $c\sub{s}^2 = \gamma p/\Sigma$.
Equations are solved on a linearly spaced polar grid ($r$, $\varphi$) with a
fixed resolution $(512, 512)$, ranging from $r\sub{min}=\SI{75}{AU}$ to
$r\sub{max}=\SI{450}{AU}$ and $\varphi \in [0,2 \pi]$. (Numerical convergence
was checked against runs with double resolution in each direction)
\MAV use finite-volumes Reimann solvers. A two-step \code{hllc} integration
scheme \citep{Harten1983} and a \code{Koren} slope limiter
\citep{Koren1993} are used in our simulations.

The model is physically inviscid. The numerical viscosity, expressed in terms of
the widely used "$\alpha$"-paradigm \citep{Shakura1973}, was estimated to lie between
$2\times10^{-8}\le$ and $\le 3\times10^{-4}$ in the vortex-forming region.
Details on this estimations are given in \cref{app:numerical_viscosity}.

The disk is truly "massless" in that both self-gravity and indirect terms due to
the barycenter's motion are neglected in the computation of the gravitational potential.
\citet{Zhu2016} showed that including either or both of these contributions affects the
vortex's evolution. In particular, the inclusion of indirect terms promotes a radial
displacement of the structure and overall increases the density contrast with respect to
its background. This latter result was also confirmed by \citet{Regaly2017} for vortices
formed in a viscosity transition region. Because of these combined effects, the velocity
of the structure is also modified, while a direct comparison is non-trivial.

\subsection{Initial conditions}
The initial gas surface density features a smooth radial density jump, modeling a disk
cavity as
\begin{equation}\label{eq:init_sigma}
  \Sigma(r, t=0) =
  \Sigma_0 \left(r/r_*\right)^{-1}
  \times \frac{1}{2} \left[1+\tanh \frac{r-r\sub{j}}{\sigma\sub{j}} \right]\:,
\end{equation}
where $r\sub{j}$ and $\sigma\sub{j}$ are the radial location and the width of the jump
respectively, and $r_*=\SI{100}{AU}$ is a normalisation factor.
The initial equilibrium azimuthal velocity is defined as
\begin{equation}\label{eq:rot_eq}
  \frac{v_\varphi^2}{r}
  = \frac{GM}{r^2} + \frac{\partial_r p}{\Sigma}\:,
\end{equation}
where $G$ is the universal gravity constant and $M$ is the mass of the central star.

Observational constraints for \hd are used to tune numerical values, wherever
applicable, as we will now detail. We assume $M = \num{2.2}$ $\Ms$, compatible with
existing estimations \citep{Verhoeff2011,Casassus2015a}.  We choose a standard
radial density slope in $r^{-1}$, which is also compatible with estimate from
\citet{Verhoeff2011} in the optically thin approximation at \SI{1}{mm}.  Distance to star
is now known with good precision $157\pm^7_6 \si{pc}$ thanks to \citet{Gaia2016}, which
implies the cavity lies at $r\sub{j}=\SI{157}{AU}$ for an angular size of
\ang[angle-symbol-over-decimal]{;;1.0} \citep{Casassus2012}. The reference setup has an
aspect-ratio, or "temperature"\footnote{The disk's vertical spreading is physically caused
by heating, and usually characterized by a scale height.} $h \equiv H(r\sub{j})/r\sub{j}
\simeq 0.09$, where $H(r)$ is the disk scale height.

\begin{figure}
  \centering
  \includegraphics[width=\figwidth]{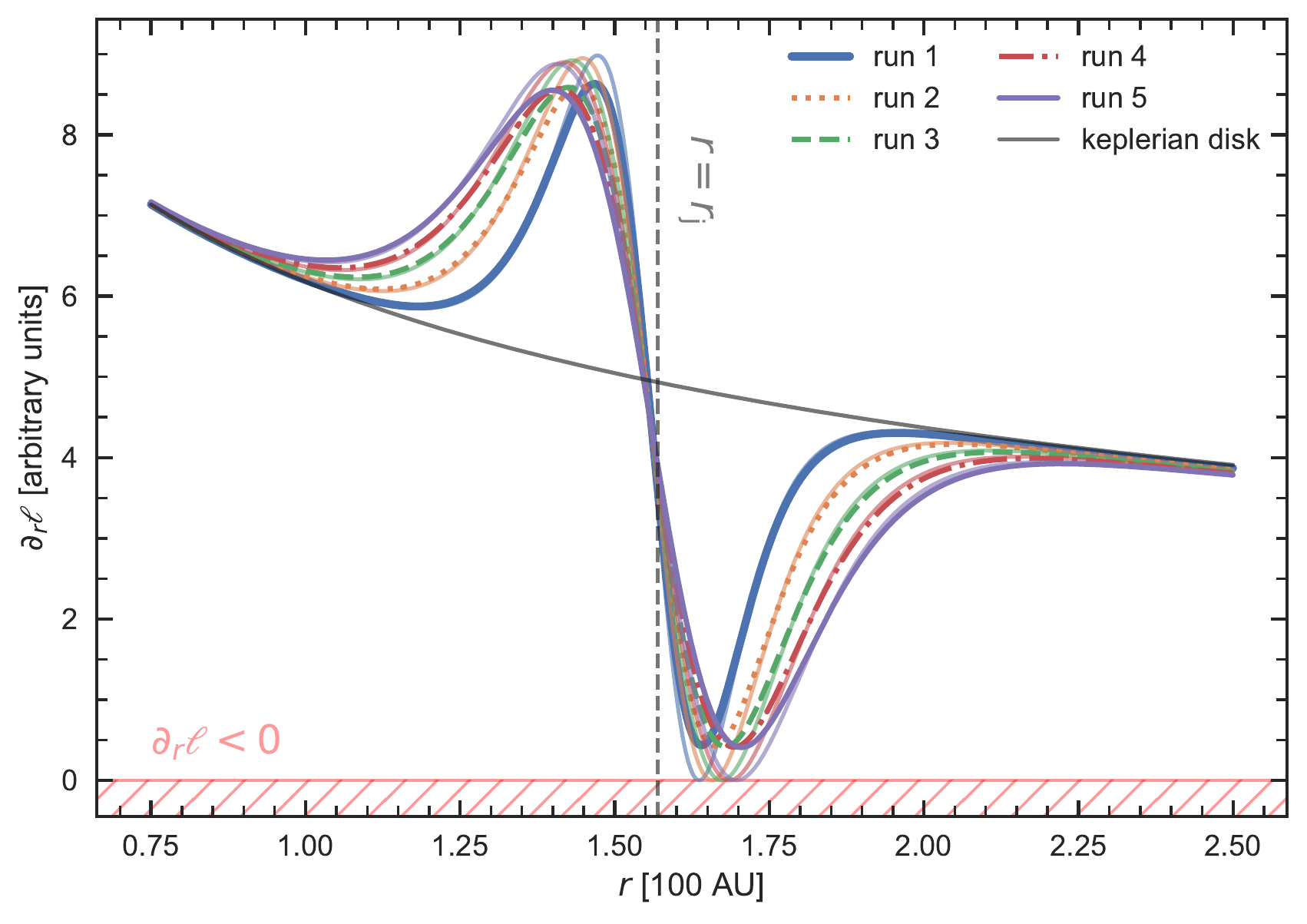}
  \caption{Initial gradient in specific angular momentum $\ell$ for our 5 simulations
  (thick lines, ranked from coldest to hottest) compared to the keplerian case. The width
  of the density ``jump'' region $\sigma\sub{j}$ is adjusted in two steps. First, we
  derive the critical value $\sigma\sub{j}^\mathrm{crit}$ for which the global minimum in
  $\partial_r \ell$ is exactly 0 (thin solid lines). Any value $\sigma\sub{j} <
  \sigma\sub{j}^\mathrm{crit}$ would give rise to a rotationaly unstable region,
  characterized by $\partial_r \ell < 0$. We then take an arbitrary \SI{5}{\%} margin and
  set $\sigma\sub{j} = 1.05\times \sigma\sub{j}^\mathrm{crit}$ in our runs (thick dashed
  lines).}
  \label{fig:regularized_specific_angular_momentum}
\end{figure}

Other simulations with $h \in [0.09;\: 0.16]$ were performed, and labeled run 1
to run 5 by increasing value in $h$. They are
discussed in \cref{ssec:detectability_vs_temperature}. The derivation of this parameter is
detailed in \cref{app:aspect_ratio}.
As this temperature is varied, we adjust the density jump's width $\sigma\sub{j}$ within
$5\%$ of its critical value, where the disk becomes rotationally unstable under Rayleigh's
criterion \citep{Rayleigh1879}. Doing so, we approach the physical upper limit in vortex
velocity after the RWI saturates. The corresponding signature in the specific angular
momentum $\ell = r v_\varphi$ is illustrated in
\cref{fig:regularized_specific_angular_momentum}. The computed values for
$\sigma\sub{j}$, and for runs from 1 to 5, are $[11.6, 14.7, 16.9, 18.7, 20.2]\;\si{AU}$.
Even in the hottest case, the simulation box extends at least $4\sigma\sub{j}$ away
from the density jump center $r\sub{j}$.
Unless explicitly stated, all figures show the results for the reference model.

\subsection{Boundary conditions}
Boundary conditions are imposed through ghost cells outside of the domain and wave killing
region in the active domain.  In the radial direction, ghost cells are fixed to the
initial equilibrium values for density and azimuthal momentum.  The radial momentum is
copied from the first cells to the ghost cells at inner boundary, and extrapolated
linearly with no-inflow condition at outer edge. However, these boundary conditions have
low impact as standard damping zones \citep{deValBorro2006} are also used to avoid
reflections at domain edges. The domain is periodic in the azimuthal direction.

\subsection{The Rossby wave instability \& vortex formation}
\begin{figure}
    \centering
    \includegraphics[width=\figwidth]{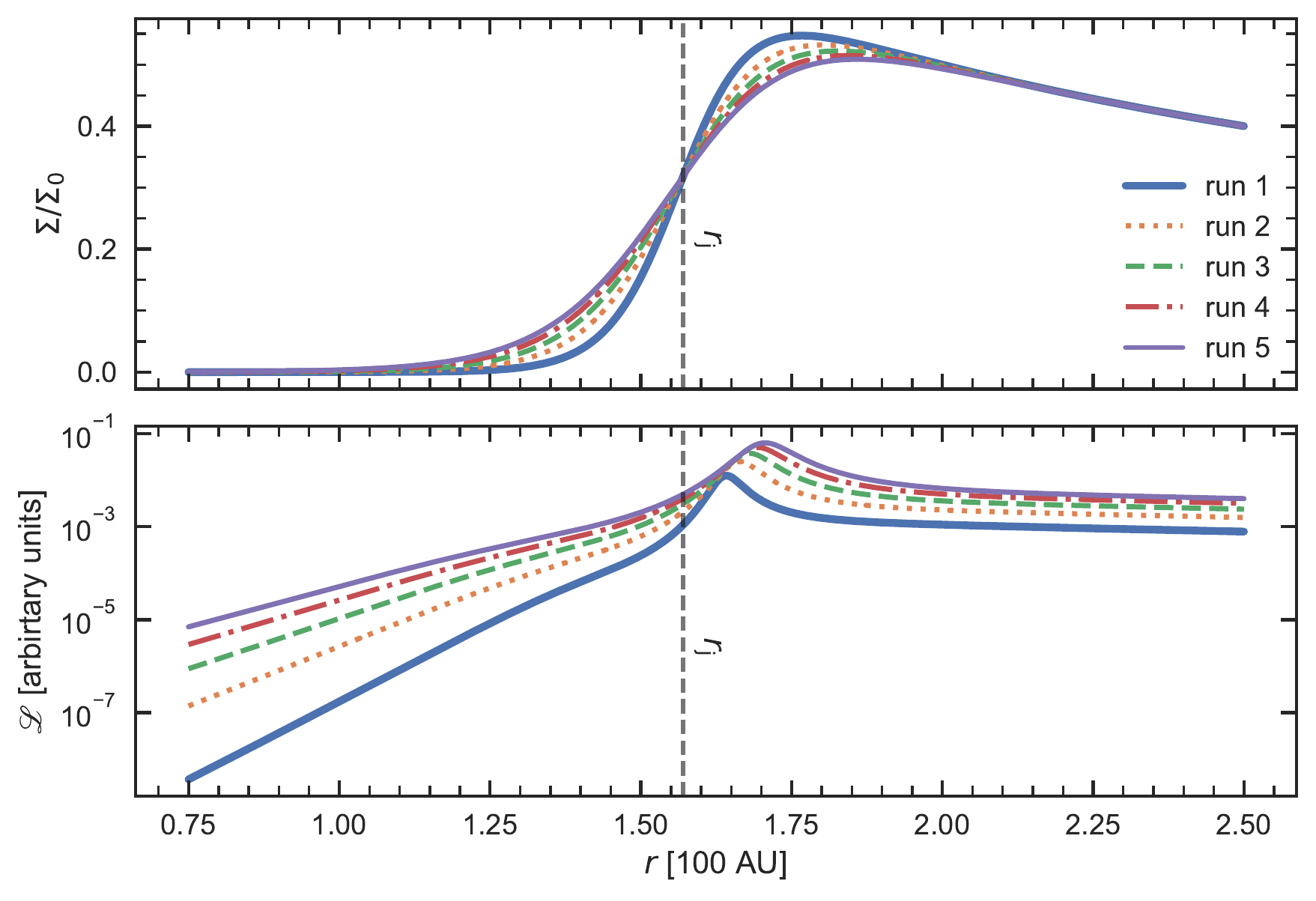}
    \caption{Initial radial profiles in surface density (top) and the
      $\mathscr{L}(r)$ function (defined in \cref{eq:lovelace}) for our 5 simulations. In
      each model, $\mathscr{L}(r)$ features a clear local maximum, which is a necessary
      condition to RWI growth.}
    \label{fig:lovelace_instable}
\end{figure}

RWI is similar to the Kelvin-Helmholtz instability in a differentially rotating keplerian
disk. It tends to convert excess shear into vorticity.  \cite{Lovelace1999} showed that a
local extremum in the background potential vorticity is a necessary condition to the RWI.
More recent works clarified that a minimum is required \citep{Lai2009, Ono2016}.  The key
function is defined as
\begin{equation}\label{eq:lovelace}
  \mathscr{L}(r)
  = \frac{1}{2}\frac{\Sigma}{\left(\Nabla\times\vec{v}\right)\cdot\vec{e_z} }S^{2/\gamma}\:.
\end{equation}
We exhibit this key function within our initial setup in \cref{fig:lovelace_instable},
showing the existence of a local maximum in $\mathscr{L}(r)$, corresponding to a minimum
in vorticity.

\begin{figure}
    \centering
    \includegraphics[width=\figwidth]{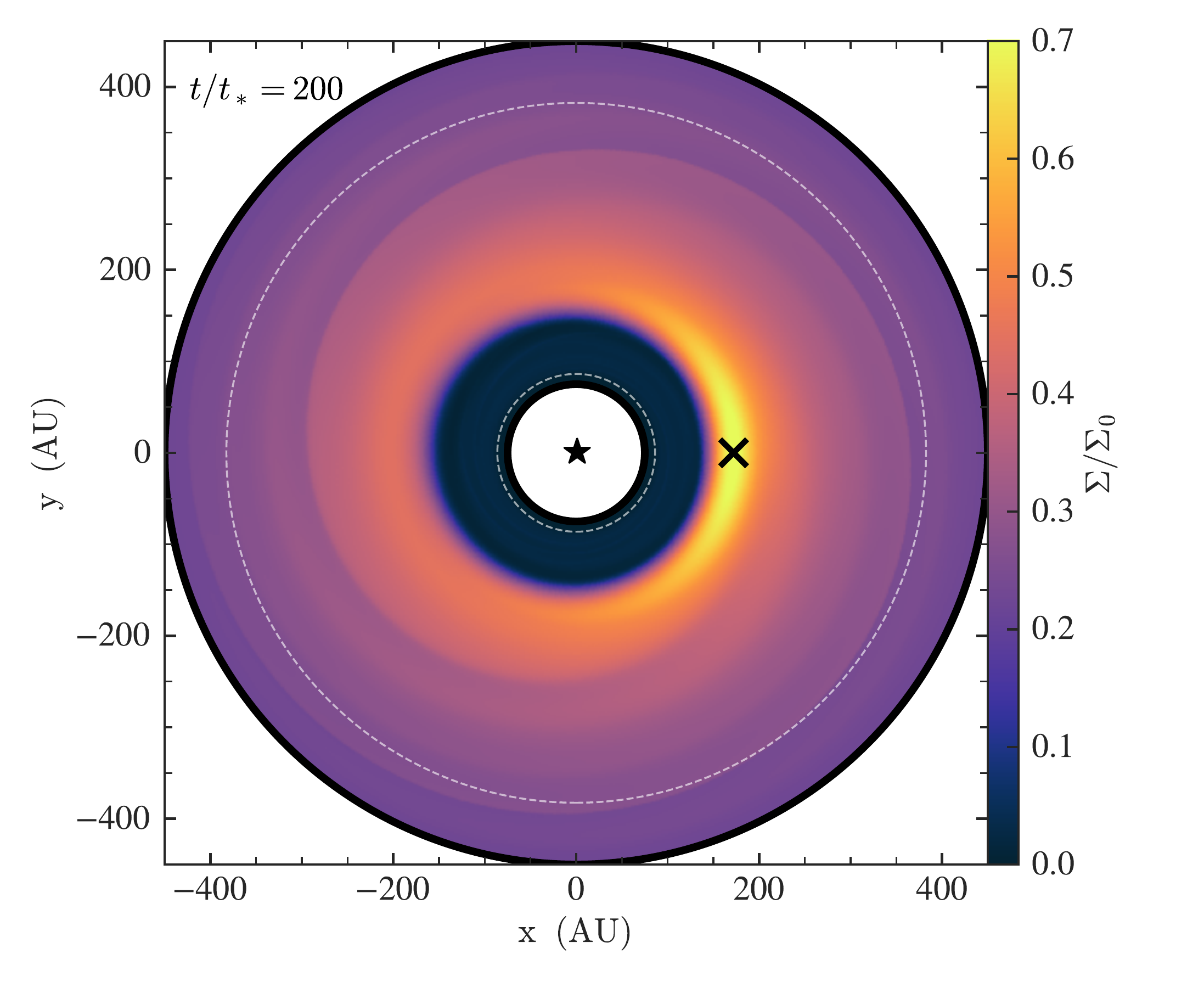}
    \caption{Gas surface density plotted in Cartesian coordinates ($x,y$) after
      $t=200$ orbital periods ($t_*$). The global density
      maximum is indicated by a black cross.  The position of the central star is marked
      as a "$\bigstar$" symbol. The simulation box radial limits are drawn as
      solid black circles, while dashed-line circles indicate the limits of wave damping
      zones.}
    \label{fig:final_state_vortex}
\end{figure}

\begin{figure}
    \centering
    \includegraphics[width=\figwidth]{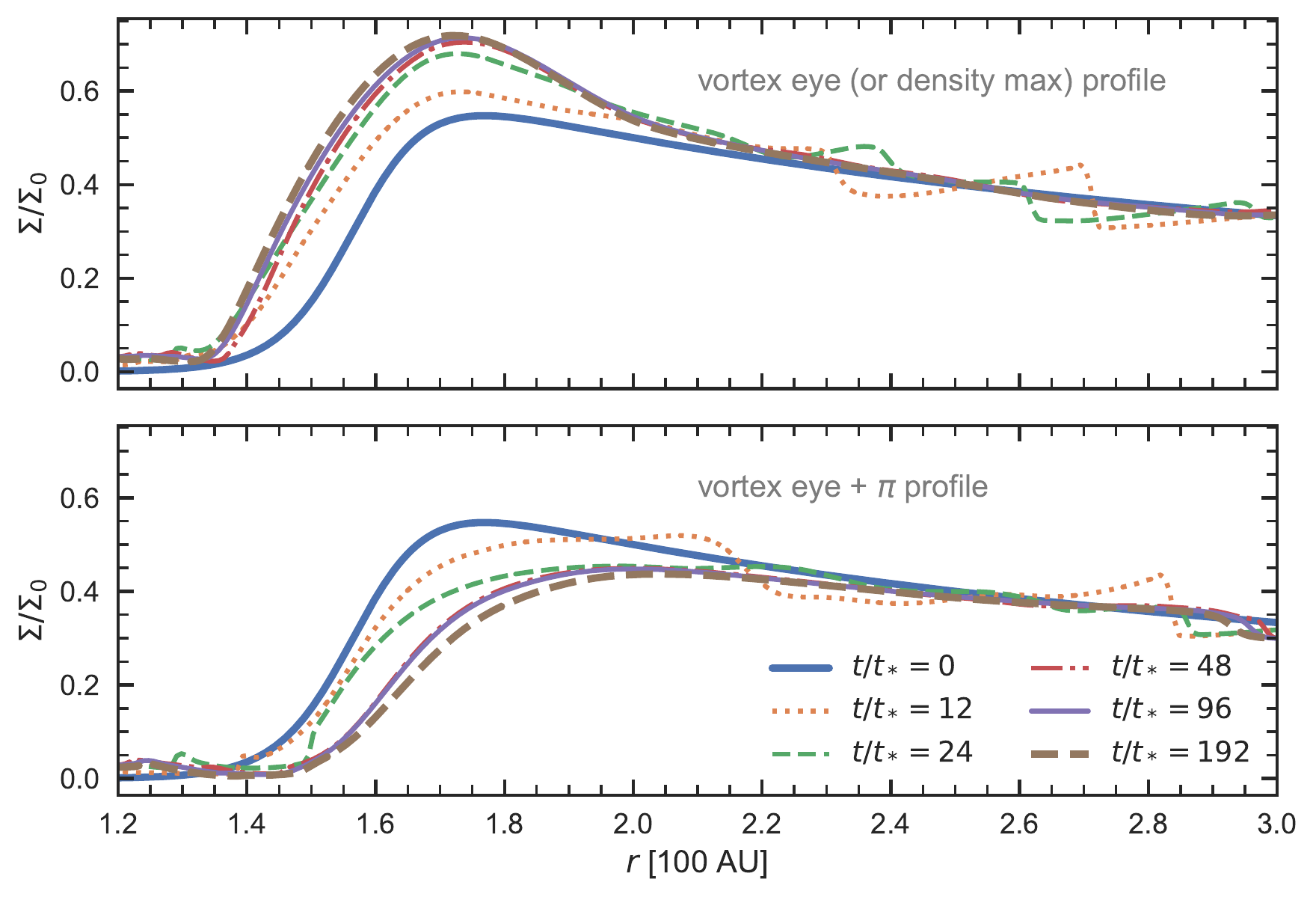}
    \caption{Time evolution of the radial density profile, plotted
            as slices at the azimuth
            of the density maximum where the vortex eye lies (top) and its radial
            opposite (bottom). The slices correspond to the $y=0$ axis in
            \cref{fig:final_state_vortex}, with $x>0$ (top) and $x<0$ (bottom)
            respectively. After $\sim 40$ orbital periods, the disk has practically
            reached a stationary state. The cavity profile itself has become uneven,
            showing a non-zero eccentricity.}
    \label{fig:conv_density}
\end{figure}

We find that, in order to excite the RWI unstable modes, it is useful to add
  random perturbations. We chose to perturb the radial velocity, which is zero otherwise,
  as
\begin{equation}
    v_r (r, \varphi, t=0) = c\sub{s} \psi(r, \varphi) \exp \frac{-\left(r - r\sub{j}\right)^2}{2\sigma\sub{j}^2}\:,
\end{equation}
where $\psi(r,\varphi) \in [-10^{-2}, 10^{-2}]$ is a uniformly distributed random variable
drawn for each grid cell. After the instability has saturated, we obtain a single vortex
shown in \cref{fig:final_state_vortex}.  In a frame co-rotating with the vortex, its
global structure is quasi-stationary as shown in \cref{fig:conv_density}.  The radial
density profile at the azimuth of the density maximum is plotted at different times.  The
orange dotted curve at $t=10$ features the most noticeable fluctuations, as smaller eddies
are still undergoing a merger, and strong spiral waves are launched outwards.  After $40$
orbital periods, the surface density of the vortex is stabilized and does not rapidly
evolve any more.  Thus we will consider this state as quasi-stationary, as we take a look
at the dynamics of the structure in the next section.

\section{Vortex signatures in dynamics}
\label{sec:signatures}
In this section, we provide observational signatures obtained from the vortex's dynamics.
The observable studied here is the velocity projected along the line of sight
$v\sub{LOS}$.  We first discuss an adequate decomposition of the velocity field to
characterize the signatures. We then study their observability against disk orientation,
and provide insight on how disk aspect ratio affects observed velocities.

\subsection{Extracting dynamical signatures}
\label{subsec:extract_dyn}

\begin{figure}
    \centering
    \includegraphics[width=\figwidth]{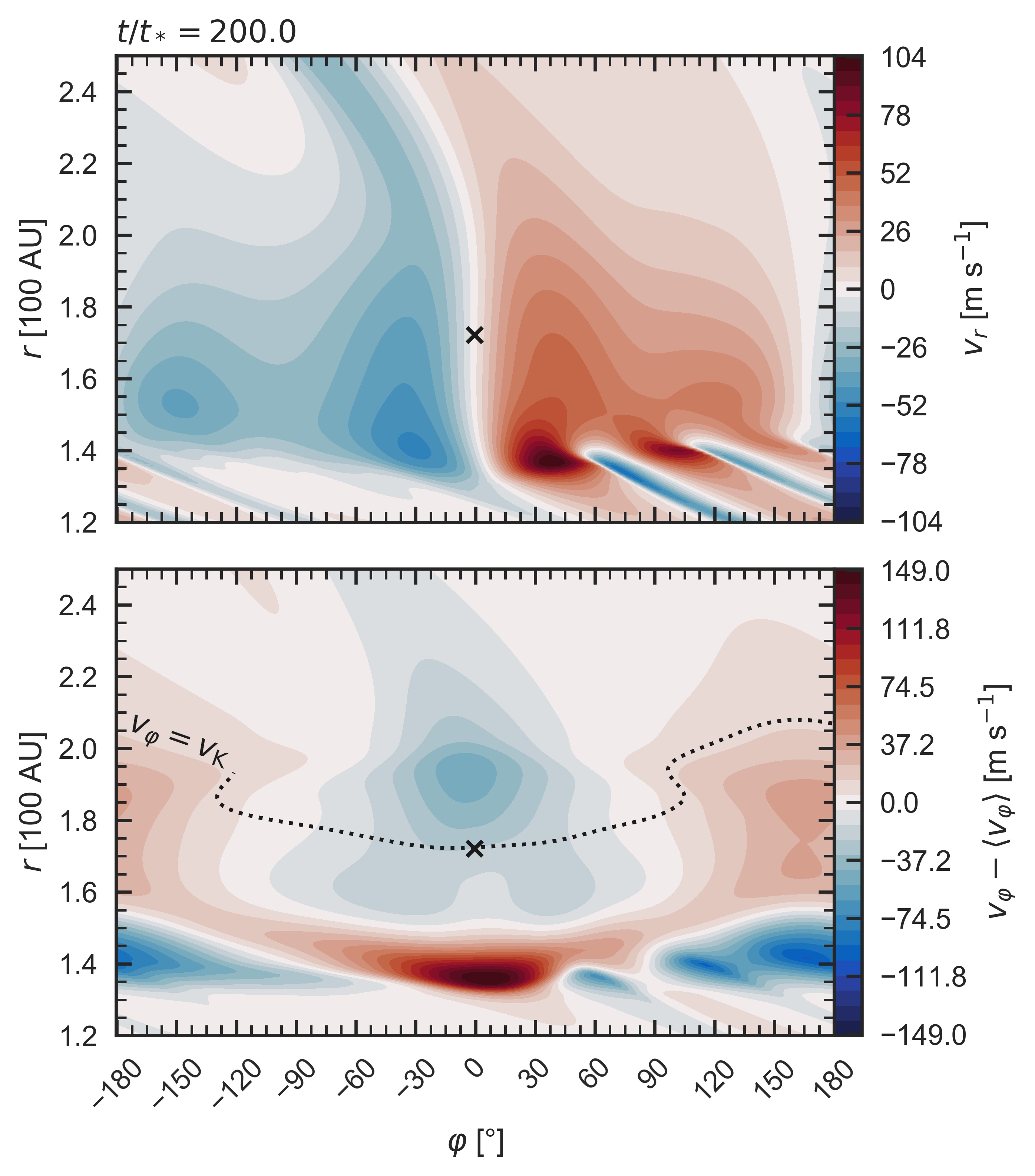}
    \caption{Polar components of a vortex's velocity field. \textit{Top\,:} radial
      velocity. \textit{Bottom\,:} azimuthal velocity, where the axisymetric part $\vsym$
      is carried out. The pressure maximum is indicated by a black cross.  The snapshot is
      taken at $t=200$ orbital periods.  A doted line in the bottom panel
      indicates fluid in exact Keplerian rotation.}
    \label{fig:velocity_res}
\end{figure}

The dynamics of a disk is dominated by rotation around the central star. In an axisymetric
stationary state, the net radial force is zero, as in \cref{eq:rot_eq}. Due to pressure
gradients, the radial equilibrium slightly departs from Keplerian motion. This is the
sub-keplerian rotation in a disk with negative radial pressure gradient. As a dynamical
structure, a vortex exposes little difference to global rotation. Thus, it is useful to
decompose the angular velocity $v_\varphi$ as
\begin{equation}
    v_\varphi = \vsym + \left(v_\varphi-\vsym\right)
    \equiv \vsym +  v_\varphi' \:,
\end{equation}
where $\langle \cdot \rangle$ is the azimuthal average operator, and we denote the
non-axisymetric part as $v_\varphi'$.
Hence the total velocity field $\vec{v}$ can be decomposed in the polar basis ($\vec{e}_r,
\vec{e}_\varphi$) as
\begin{equation}\label{eq:vel_decomp}
    \vec{v} = v_r \vec{e}_r + \left( \vsym + v_\varphi'\right) \vec{e}_\varphi
    \equiv \vsym \vec{e}_\varphi + \vec{v'} \:,
\end{equation}
In the absence of a global accretion flow, there is no relevant axisymetric part in $v_r$.
Hence we consider that dynamical signatures of non-axisymetric features reside in
$\vec{v'} = v_r \vec{e}_r+ v_\varphi' \vec{e}_\varphi$. Both components of this residual
radial and azimuthal velocity are quantified in \cref{fig:velocity_res}, and are of
similar amplitudes. For comparison, the typical Keplerian speed at the vortex position
($r\sim \SI{180}{AU}$) is $v\sub{K} = \SI{3.3}{km.s^{-1}}$, one to two orders of
magnitudes larger than the deviation due to the vortex, and one order of magnitude larger
than the local sound-speed $c\sub{s}$. The amplitude in the azimuthal velocity is
as high as \SI{300}{m.s^{-1}} for this reference (coldest) model. This sets a first upper
limit to the spectral resolution required for a direct detection to about
\SI{100}{m.s^{-1}}. This is achievable for bright lines with ALMA, e.g., for the
CO (2-1) or the CO (3-2) transitions by using channel widths of \SI{70}{kHz} or
\SI{120}{kHz} (or narrower), respectively. For example, \citet{VanDerMarel2016a}
successfully detected the \chem{^{13}CO} (3-2) and \chem{C^{18}O} (3-2) lines of SR21, HD
135344B, DoAr44, and IRS 48 with good signal-to-noise (SNR) ratio (peak SNR in the
integrated intensity map up to 30 for the \chem{^{13}CO} line)  with spectral resolution
of \SI{0.1}{km.s^{-1}} and angular resolution of 0\farcs25. \citet{Boehler2017} obtained
data with similar angular and spectral resolutions for \hd but with much higher SNR. All
sources are well detected in the lines and increasing the spectral resolution by another
factor of 2, as well as the SNR, is possible within a reasonable amount of time
($<\SI{12}{hours}$). We note that, there is a non-zero azimuthal velocity deviation at
the maximum density/pressure (i.e. $v_\varphi'\neq 0$), as seen in
\cref{fig:velocity_res}. Indeed, the vortex being an asymmetric structure, the radial
position of the pressure extremum varies with the azimuth. Consequently, the line of exact
keplerian rotation is not circular as shown in \cref{fig:velocity_res}.

\begin{figure}
    \centering
    \includegraphics[width=\figwidth]{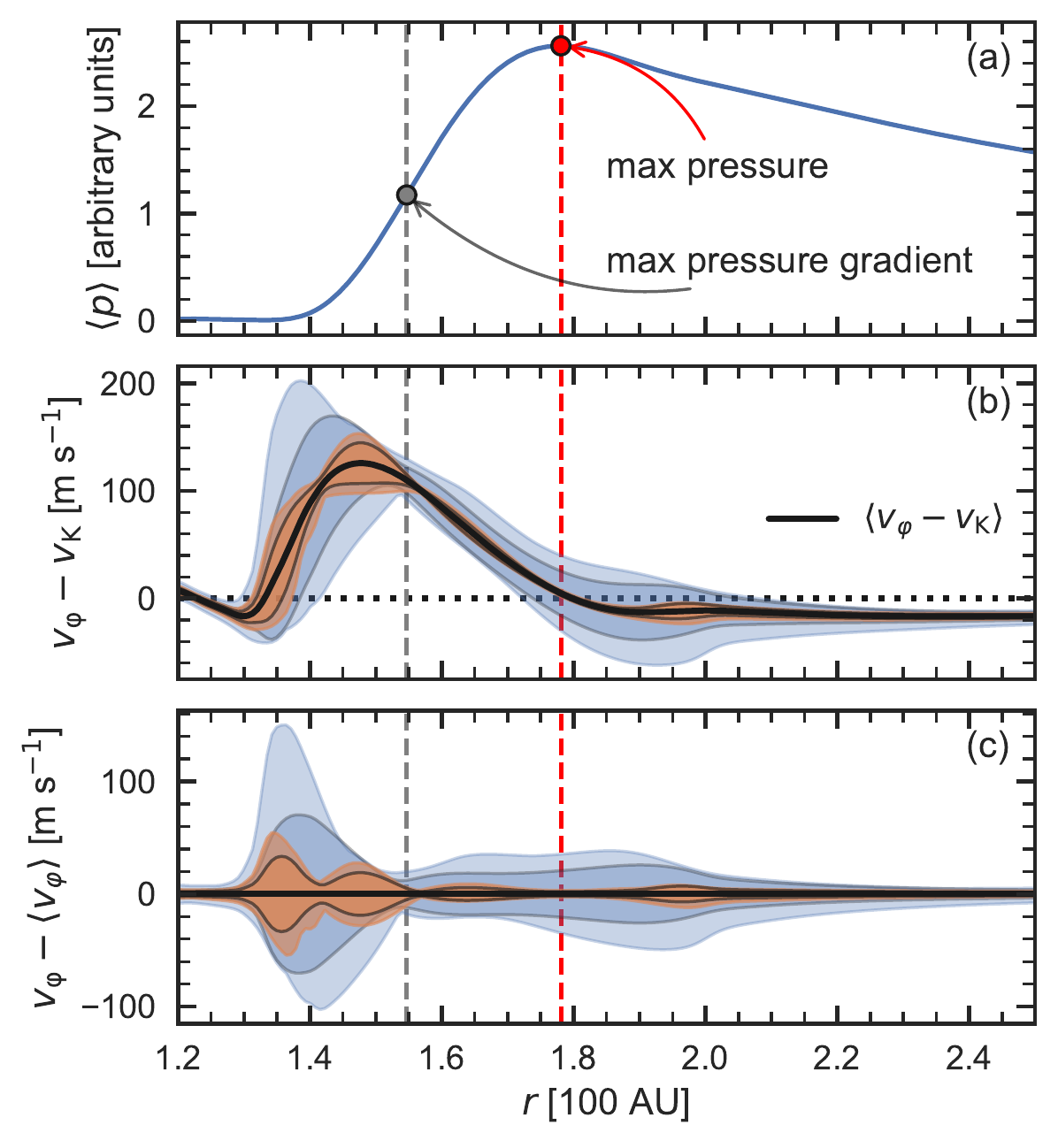}
    \caption[]{A comparison between Keplerian velocity $v\sub{K}$ and
    average azimuthal velocity $\langle v_\varphi \rangle$ as 1D masks.
    \textbf{(a)}: pressure profile in arbitrary units.
    \textbf{(b and c)}: azimuthal velocity cross-sections, with offsets (masks)
    $v\sub{K}$ and $\langle v_\varphi \rangle$
    respectively. The average $\langle V - \mathrm{mask}
    \rangle$ is indicated by a thick black line, while shadows show the
    amplitude and standard deviations in blue ($V=v_\varphi$), and orange  ($V =
    \frac{1}{2} \left(v_{\varphi, \mathrm{left}} + v_{\varphi,
    \mathrm{right}}\right)$). Data is taken at $t=200$ orbital
    periods.\footnotemark
    }
    \label{fig:pressure_vs_vrot}
\end{figure}

However, the decomposition proposed in \cref{eq:vel_decomp} is vain unless the proposed
dominant term $\vsym$ can be subtracted from observations.  While a Keplerian fit is
usually a suiting approximation of the dominant velocity term, it proves insufficient near
sharp density jumps.  As shown in \cref{fig:pressure_vs_vrot} (a, b), subtracting a
Keplerian power law leaves systematic velocities caused by pressure gradients.
In the density transition region, those systematics dominate over the variability
in the remaining signal.

However, we further show (\cref{fig:pressure_vs_vrot} (b, c)) that averaging two facing
cross sections in azimuthal velocities consistently yields a much better approximation for
the global azimuthal average $\langle v_\varphi\rangle$, with a standard
deviation $\le \SI{20}{m.s^{-1}}$. This is a direct sign that the
non-axisymetric parts of the azimuthal velocities $v_\varphi'$ in opposing disk halves are
anticorrelated, although not strictly equal in amplitudes.  Given that on the disk's
observational major axis $x$, detection is only sensitive to azimuthal velocities, we
naturally obtain a satisfying method to subtract $\langle v_\varphi\rangle$ from the whole
signal.  Consequently, we will now confidently assume that the axisymetric part $\vsym$ of
the azimuthal velocity can indeed be removed with good precision from observations, and
will only consider the remaining components of $\vec{v'}$ exhibited in
\cref{fig:velocity_res}.

\footnotetext{Inspiration for this figure was drawn from \citet{Teague2018a}.}

\subsection{Vortex detection in line-of-sight velocities}
\label{sec:detection_los_vel}
\begin{figure}
    \centering
    \includegraphics[width=0.8\figwidth]{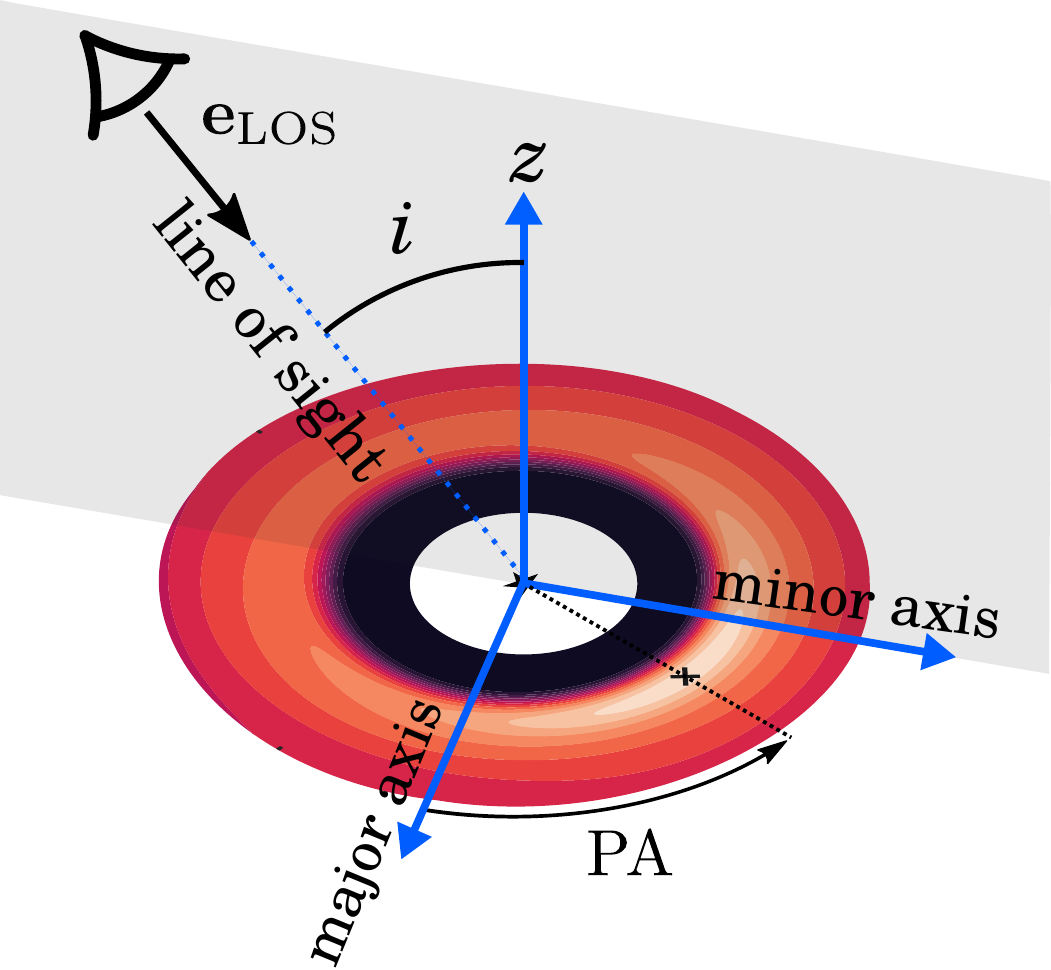}
    \caption{A sketch view of our notations. A grey shadow shows the plane containing the
    vertical $\vec{e_z}$ and the line of sight.}
    \label{fig:schema_orientation}
\end{figure}

Gas velocity is usually detected through Doppler-shifting in molecular lines.  It is
therefore the velocity component parallel to the line of sight that is probed.  Within
optically thick lines, the resulting velocity profile can be blurred by vertical
integration over disk height.  It is beyond the scope of the present work to inquire on
this second-order effect, so we neglect both optical and geometrical thickness effects.
This approach is reasonable within the assumption that emissive molecular regions are
geometrically thin and well resolved (as remarked by \citet{Teague2018b}). Furthermore,
full 3D simulations showed that, in a stationary state, a vortex tends to be tubular and
that its vertical velocity is negligible \citep{Lin2012, Zhu2014, Richard2013}.  This
comforts us in the idea that, for long lived vortices, it is reasonable to ignore this
component\footnote{This also means we ignore the vertical extension in the conical shape
of the emissive layer. However, it can easily be shown that for inclinations lowers than
\SI{45}{\degr}, even a very high emissive layer $z\simeq 5H$ and a large aspect ratio
$H/r=0.2$, can in principle be deprojected as long as it remains spatially thin}.  To
study vortex dynamical signatures, we use here cylindrical coordinates centered on the
star. The radial axis ($\varphi = 0$) is the observational major-axis, and the upper part
of the disk ($z>0$) is defined to be the one seen by the observer
\cref{fig:schema_orientation}.

\begin{figure*}
    \centering
    \includegraphics[width=\textwidth]{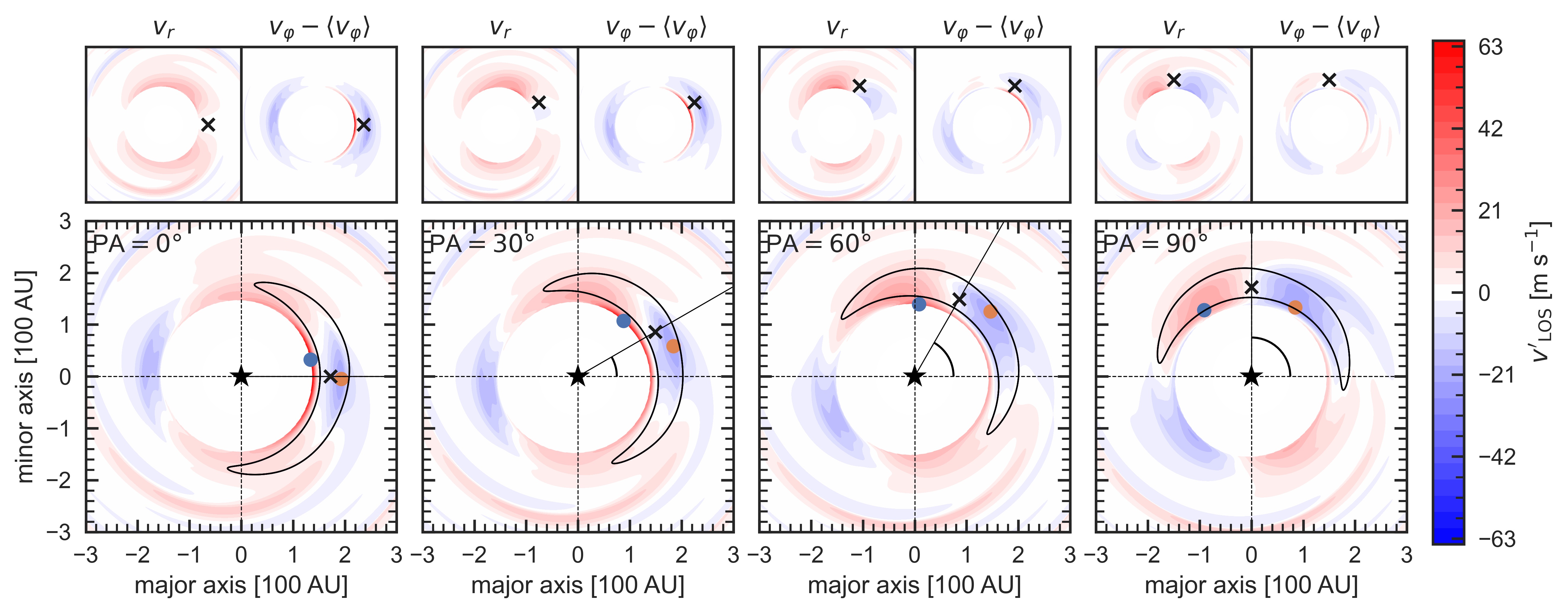}
    \caption{Line-of-sight velocities (bottom large panels) as defined in \cref{eq:vproj},
      applied to \hd with $i=\SI{27}{\degr}$, and varying PA. Top panels exhibit the
      corresponding polar components. Color reflects implied Doppler-shifts in molecular
      lines.  Blue/orange dots indicate extreme values in $v'\sub{LOS}$.  In the leftmost
      panel, the vortex's spatial extension is shown as a solid contour which corresponds
      to $\Sigma = 0.5 \Sigma_0$, where $\Sigma_0$ is the scaling factor used in
      \cref{eq:init_sigma}.
      The inner cavity, where fast spiral waves are launched but surface density is low,
      is not shown here\,: regions with $\Sigma/\Sigma_0 < 0.1$ are masked. As a
      proxy for the vortex's eye, a black cross indicates the density maximum.}
    \label{fig:proj_ideal}
\end{figure*}

Thus, the line-of-sight direction $\vec{e\sub{LOS}}(i, \varphi)$, defined as pointing away
from the observer, can be written in the disk cylindrical basis $(\vec{e}_r,
\vec{e}_\varphi, \vec{e}_z)$ as
\begin{equation}
    \vec{e\sub{LOS}}(i, \varphi) =  -\left(\sin(i) \cos(\varphi) \vec{e}_r
                                + \sin(i) \sin(\varphi) \vec{e}_\varphi
                                + \cos(i) \vec{e}_z\right)\:,
\end{equation}
and it follows that the line-of-sight velocity corresponding to $\vec{v'}$ writes
\begin{equation}\label{eq:vproj}
  v\sub{LOS}' \equiv \vec{v'}\cdot\vec{e\sub{LOS}}
  = -\sin(i) \left(\sin(\varphi) v_r + \cos(\varphi) v_\varphi'\right)
\:.
\end{equation}
Hence, the effective observable $v\sub{LOS}'$ mixes $v_r$ and $v_\varphi$.  For a 2D
vortex disk inclination equally affects all projected velocities and only acts as a
scaling factor $\sin(i)$. An inclination $i=\SI{27}{\degr}$, corresponding to
the estimated value for \hd \citep{Fukagawa2013}, is used in the following applications.
This choice is arbitrary and used as a textbook case. We note that this
inclination is moderate. Deprojection would still be feasible at up to $i=\SI{45}{\degr}$,
where projected velocities would be 1.5 times larger, making detection easier.
\Cref{fig:proj_ideal} shows the morphology of the observable $v'\sub{LOS}$ (large panels),
along with corresponding components $v_r$ and $v'_\varphi$ (small panels), for four
different values of PA.  This result constitutes an idealized case, built on the
assumption that the axisymetric component $\vsym$ can be exactly subtracted from
observational data. At all position angles (PAs), the vortex's anticyclonic motion around
the density maximum is apparent in $v'\sub{LOS}$\footnote{We are set in the particular
case where the PA "rotates" in the same direction as the disk. When not so, sign in
$v'\sub{LOS}$ must simply be inverted.}. This point roughly coincides with the maximum
luminosity at most wavelengths, and can be located within continuum observations, if not
directly in molecular lines used to infer projected velocities.

We note that the vortex's eye and the region immediately facing it have similar
Doppler shifts (e.g. both blue at $\mathrm{PA}=\SI{0}{\degree}$). This is an expected
outcome of subtracting the azimuthally averaged velocity, since the both regions are local
extrema along the azimuthal direction. Another signature of the vortex is the azimuthal
proximity between the maximum density (black cross) and the projected velocity extrema
(color dots). The later two points are determined by the physical on-site
velocity as well as the system's inclination, and hence are virtual positions. Their
physical separation is maximized for $\mathrm{PA}=\SI{90}{\degr}$ and minimized for
$\mathrm{PA}=\SI{0}{\degr}$. A direct implication is that detecting a vortex lying on the
major axis requires greater angular resolution. However, little dependency of
the velocity range on the PA is found. The topography of the signal changes with the PA
but the anticyclonic region stands out regardless the orientation. The
signature is also typical with a sign reversal in the vicinity of the pressure maximum,
along the major-axis direction. This characteristic behaviour, sign change, is
easier to measure in relative than the absolute small velocities and would be observed
even with a beam covering the vortex almost entirely (about \SI{100}{AU}, or
\ang[angle-symbol-over-decimal]{;;0.6} in the case of \hd).

\subsection{Detectability against disk temperature}
\label{ssec:detectability_vs_temperature}
\begin{figure*}
    \centering
    \includegraphics[width=\textwidth]{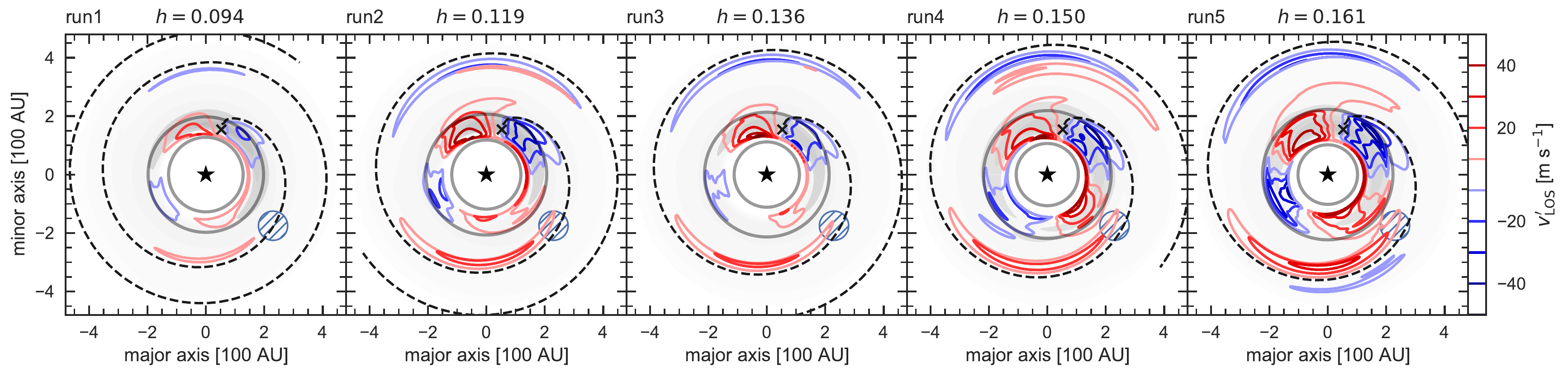}
    \caption{A comparative view of $v\sub{LOS}'$ with varying disk temperature $h$. Disk
      orientation is taken consistent with values found in the literature on \hd
      ($i=\SI{27}{\degr}$ \citet{Fukagawa2013}, $\PA = \SI{71}{\degr}$
      \citet{Kataoka2016}). Here we mock a spectral resolution of \SI{12.5}{m.s^{-1}}.
      The reference setup occupies the leftmost panel. Velocities in the cavity are masked
      as in \cref{fig:proj_ideal}.  In dashed lines, we over-plot the best fit spirals
      following \citet{Huang2019} (eq. 2 therein), based on linear perturbation theory
      \citep{Goldreich1979, Rafikov2002, Muto2012}. Those fits were computed using $v_r=0$
      contours as input data.  As a visual indicator, surface density is
      underplotted in greyscale. Additionally, grey circles indicate the
      $3\sigma\sub{j}$ region around the vortex eye, which is used later in
      \cref{fig:velocity_contrast_hd142}. As in other figures, a black cross indicates the
      density maximum.}
    \label{fig:vproj_hd142_sprials}
\end{figure*}

Although our setup is constrained by observations, its temperature (or equivalently $h$)
is not. Indeed the temperature gives the sound-speed, which is key in estimating the
vortex velocity. In order to study this dependency, four additional simulations with
higher temperatures ($h \in [0.094, 0.119, 0.136, 0.150, 0.161] $) were
performed.  In \cref{fig:vproj_hd142_sprials}, we show contours of projected velocity
$v'\sub{LOS}$, sampled at an interval corresponding to a tenth of the obtained
dynamical range, namely 10 \si{m.s^{-1}}. The reference, "coldest"
setup produces the lowest velocities ranging from -20 to +20 \si{m.s^{-1}}, where most of
the "detected" structure is within the vortex region.
The direct observation of a peak-to-valley velocity shift of about
\SI{40}{m.s^{-1}} is challenging but is within reach of ALMA. \citet{Boehler2017} observed
\hd for a total of \SI{4}{hours} during Cycle 1, targeting the continuum and
\chem{^{13}CO} (3-2) and \chem{C^{18}O} (3-2) lines with a spectral resolution of
\SI{110}{m.s^{-1}} (after Hanning smoothing). The disk is detected in both lines at high
SNR. The angular resolution of the observations was \SI{45}{AU} (beam $0\farcs27 \times
0\farcs31$. The presence of a velocity signature is currently being investigated in that
data set (Boehler et al, in prep.). This angular resolution is sufficient to resolve the
vortex in \hd and the spectral resolution can be improved by a factor of 2 on the brighter
\chem{^{12}CO} line \citep{Perez2015}, or on the
\chem{^{13}CO} and \chem{C^{18}O} lines by increasing the time spent on-source.

\begin{figure*}
  \centering
  \includegraphics[width=\textwidth]{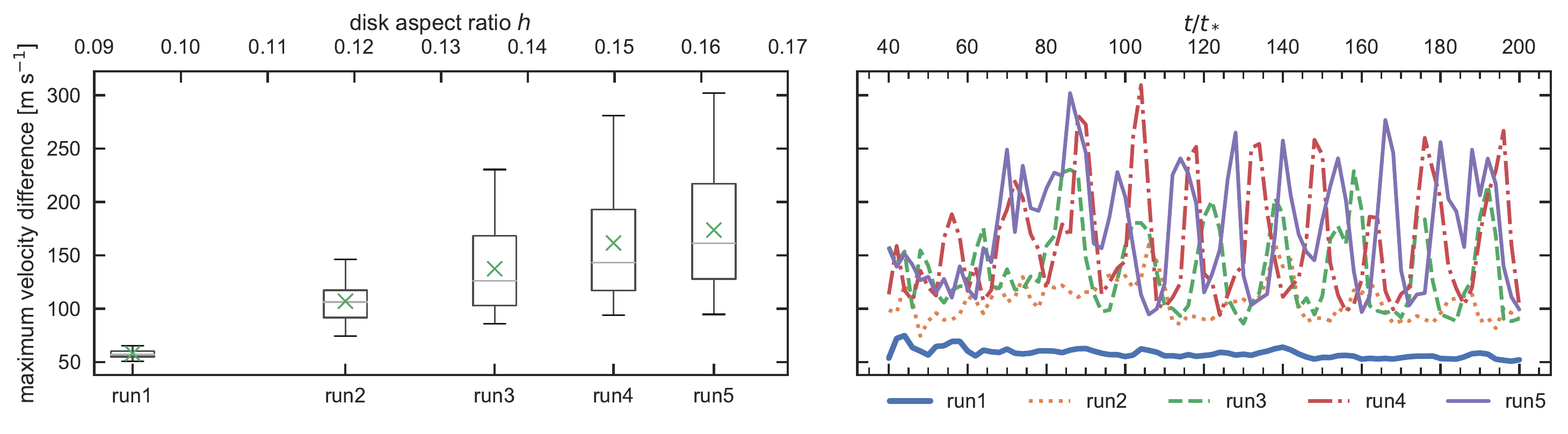}
  \caption{Amplitude in $v'\sub{LOS}$ across the annular region shown in
    \cref{fig:vproj_hd142_sprials}, throughout the simulation time, represented
    as a boxplot (left). The whole time series is unraveled in the right panel. It is
    sampled every 2 orbital periods (the output rate of our simulations). Although the
    first run is remarkably constant, runs 3 to 5 exhibit significant dispersion within
    this metric, while the mean value (green crosses) itself is stabilized. These
    oscillations' period corresponds to the life-cycle of a secondary spiral arm,
    illustrated in
    \cref{fig:transient_secondary_spiral}.}
  \label{fig:velocity_contrast_hd142}
\end{figure*}

\begin{figure*}
\centering
\includegraphics[width=\textwidth]{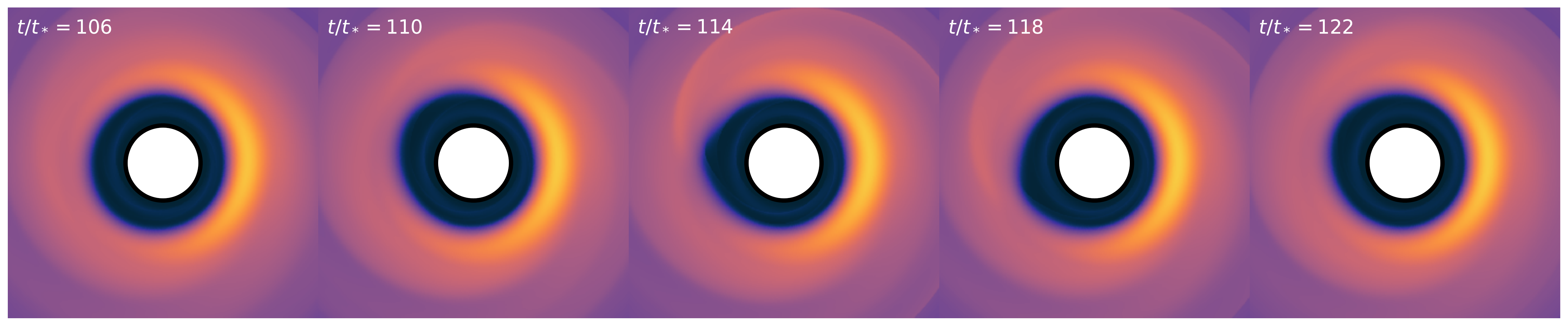}
\caption{Formation/dissipation cycle of a secondary spiral arm connected with
disk eccentricity, illustrated for the most prominent case, run 5. Color maps density
(same scale as \cref{fig:final_state_vortex}). This secondary spiral is a transient and
periodic phenomenon, responsible for large oscillations in maximum projected velocity as
measured in
\cref{fig:velocity_contrast_hd142}}
\label{fig:transient_secondary_spiral}
\end{figure*}

At higher
temperatures, more structure is revealed as the spiral arm unravels.
\Cref{fig:velocity_contrast_hd142} shows the $v'\sub{LOS}$ variation amplitude,
against temperature (left panel) and time (right panel). Although the upper bound of this
range consistently increases with temperature, we note that the mean value is almost
unchanged from run 4 to run5. Indeed, in runs 3 to 5, the amplitude of time-variations are
significantly higher than for the reference run. These large variations are related with
the life-cycle of a secondary spiral arm that appears in hot cases, as illustrated in
\cref{fig:transient_secondary_spiral}.
However, because this secondary spiral is most
prominent when the disk itself becomes visibly eccentric it is likely that this structure
would be affected, be the indirect gravitational terms taken included in the model.
A conservative conclusion is that only the lower boundary of the variation interval
should be taken into account. Additionally, we observe that between runs 4 and 5, the
dynamical range stagnates at $\SI{94}{m.s^{-1}}$\footnotemark. This saturation is likely
caused by the instability in the cavity's eccentricity, thus we infer that validity of
massless disk models is disputable in the hottest case (run 5). We note that a
\footnotetext{Indeed, this is the range shown in \cref{fig:vproj_hd142_sprials}, where the
simulations are shown at a time that minimizes it ($t/t_* = 200$).} previous study of the
gas dynamics in \hd \citep{Casassus2015a} did not provide evidence of any strong
asymmetric structure, this may indeed be due to a lack of spectral resolution ($\sim
\SI{1}{km.s^{-1}}$).

\subsection{RWI spirals}
\label{ssec:rwi_spirals}
Spirals structures are detected in \hd \citep{Fukagawa2006, Casassus2012, Rameau2012,
  Avenhaus2014, Christiaens2014}. Several scenario have been proposed to understand their
  origin, such as self-gravitational instability (SGI), excitation by the stellar
  companion \citep{Biller2012, Price2018}, connection to a shadow cast by a misaligned
  inner disk \citep{Montesinos2016}, or a combination of several effects
  \citep{Christiaens2014}. Spirals are also a natural outcome of the RWI, as Rossby waves
  are coupled to spiral density waves in a Keplerian disk. Such spiral waves would have
  the same frequency as the Rossby wave creating the vortex.

As opposed to companion-excited spirals, those are not caused by gravitational interaction
and are observed in massless disks simulations such as ours \citep{Huang2019}. For Rossby
vortices, the launching point is radially close to the vorticity extremum, and the spiral
co-rotates with the vortex. As a consequence, for spiral arms with different launching
points, the RWI explanation may be safely rejected.

However, it must be noted that the apparent launching point of the spiral, i.e.
the origin of its detectable part, graphically indicated as a blue hatched mark, will
depart from its physical origin, namely the vortex's eye. For instance,
\cref{fig:vproj_hd142_sprials} shows a $\sim \SI{90}{\degree}$ discrepancy between the
actual launching point and the the apparent origin of the main spiral arm. The figure also
shows that, considering only spectral resolution as an experimental limitation, plane-RWI
spirals are detectable as soon as the sensitivity is sufficient to resolve the vortex's
bulk signature. In short, spirals produce projected velocities just marginally smaller
than the vortex's bulk. We further note that plane-RWI spirals are a pure tracer of radial
velocities $v_r$, which are observationally characterized by a change of sign in projected
velocities across the major axis.

\begin{figure*}
    \centering
    \includegraphics[width=\textwidth]{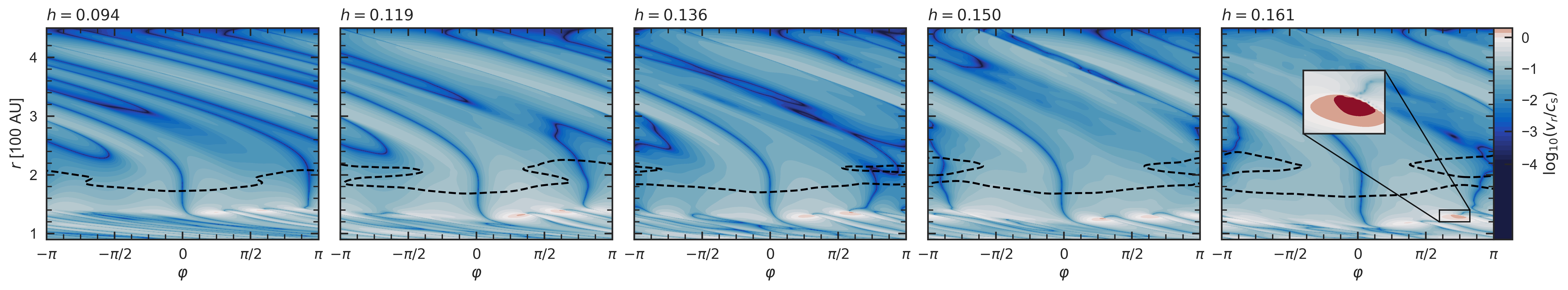}
    \caption{Radial Mach number VS temperature ($h$), seen in polar coordinates.  The
      colormaping is such that sub(super)-sonic regions appear in blue (red).  The vortex
      center is always located at $\varphi=0$.  Dashed black lines indicate $v_\varphi =
      v_\mathrm{K}$.  The global structure is not self-similar when $h$ varies, as one can
      see the keplerian line undergoes a reconnection as temperature increases,
      and spirals in the outer disk are shocking (Mach 1, white) closer to the vortex.
      Small super-sonic (red) regions are found in the inner region of the disk 
      ($r\simeq \SI{120}{AU}$), as is highlighted in an inset in the rightmost panel. The
      flow remains sub-sonic everywhere else.
      }
    \label{fig:radial_mach_number}
\end{figure*}

We note that the spiral's pitch angle increases with $h$, as a consequence of a
higher sound-speed.
Hence, radial velocities are not self-similar across our models, as hotter disks produce
higher Mach numbers, as illustrated in \cref{fig:radial_mach_number}.

\section{Discussion}\label{sec:discussion}

\subsection{Numerical VS practical differences}
\begin{figure*}
    \centering
    \includegraphics[width=\textwidth]{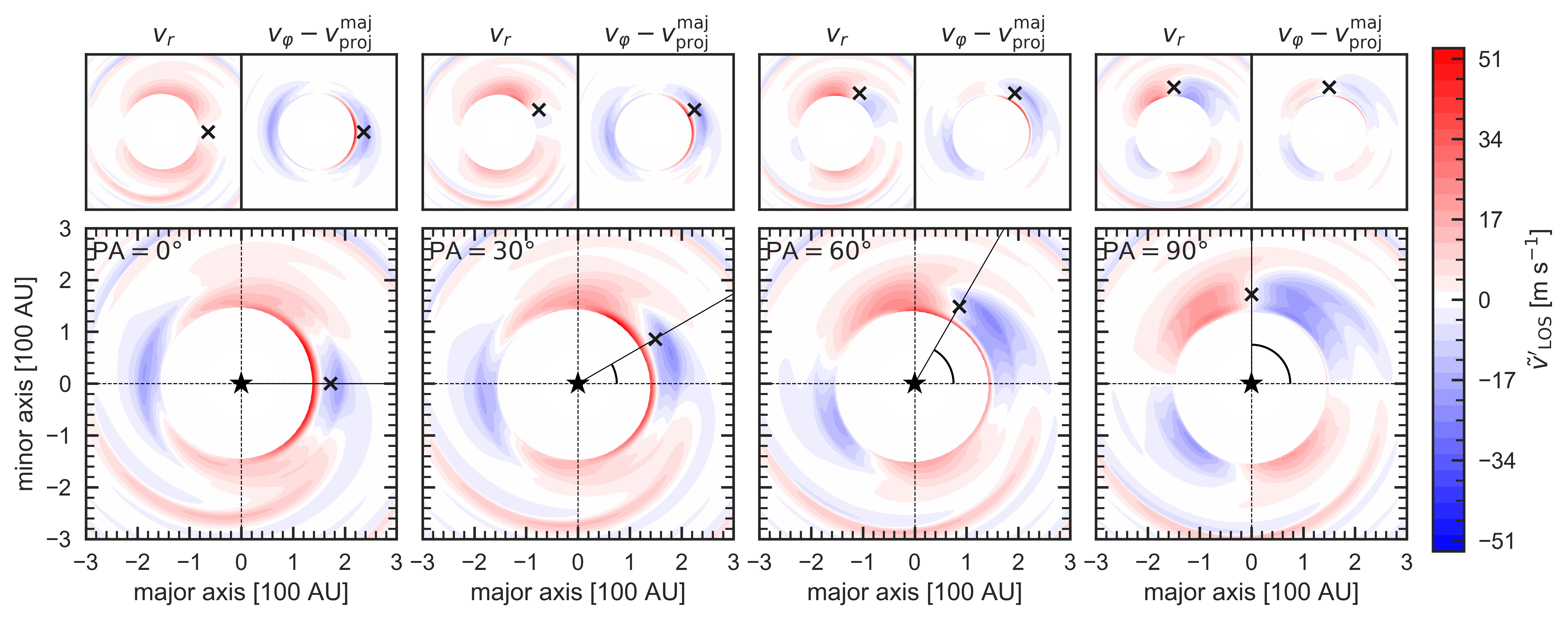}
    \caption{A practical application of our data reduction method. The figure is similar
      to \cref{fig:proj_ideal} except that $v\sub{proj}^\mathrm{maj}$ is subtracted
      instead of $\langle v_\varphi \rangle$.}
    \label{fig:proj_real}
\end{figure*}

\begin{figure*}
  \centering
  \includegraphics[width=\textwidth]{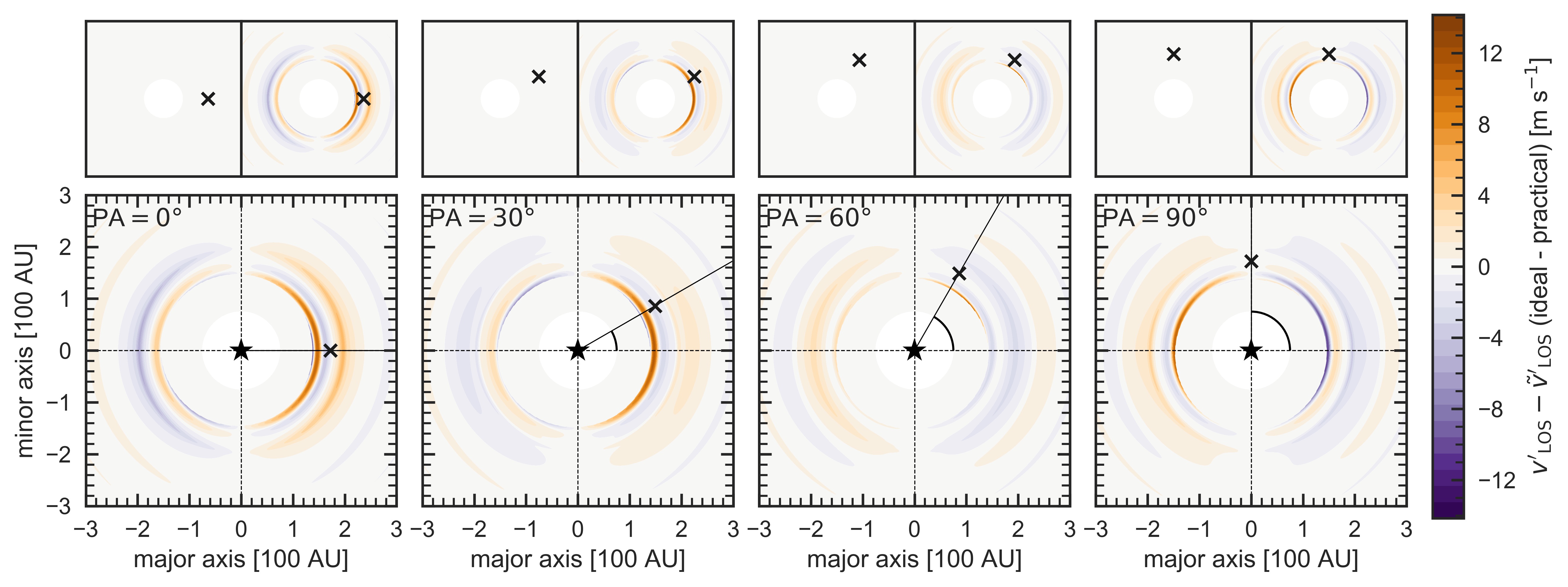}
    \caption{Difference between numerical (\cref{fig:proj_ideal}) and practical
      (\cref{fig:proj_real}) cases. By construction, $v'_\mathrm{LOS} -
      \tilde{v}'_\mathrm{LOS}$ is a separable function $\mathrm{err}(r, \varphi) = f(r)
      \cos(\varphi)$ where the density mask is axisymetric.}
    \label{fig:discrepancy_ideal_practical}
\end{figure*}

In \cref{subsec:extract_dyn}, we showed that a promising data reduction strategy for
vortex dynamic extraction in sharp density jumps was to subtract $\vsym$, and that the
projected velocity seen on the major axis ($v\sub{proj}^\mathrm{maj}$ for shorts) gives a
reasonable proxy for it.  In order to test the error implied by this approximation, this
strategy is applied in \cref{fig:proj_real}. Consistently with our previous estimation,
this more realistic view shows very little difference to the first, idealized one
(\cref{fig:proj_ideal}).  \Cref{fig:discrepancy_ideal_practical} quantifies that 2D
discrepancy as a difference between the numerical and practical cases.  We find the
discrepancy to reach at most $\sim \SI{7}{m.s^{-1}}$.

\subsection{Spiral detection}
\begin{figure}
    \centering
    \includegraphics[width=\figwidth]{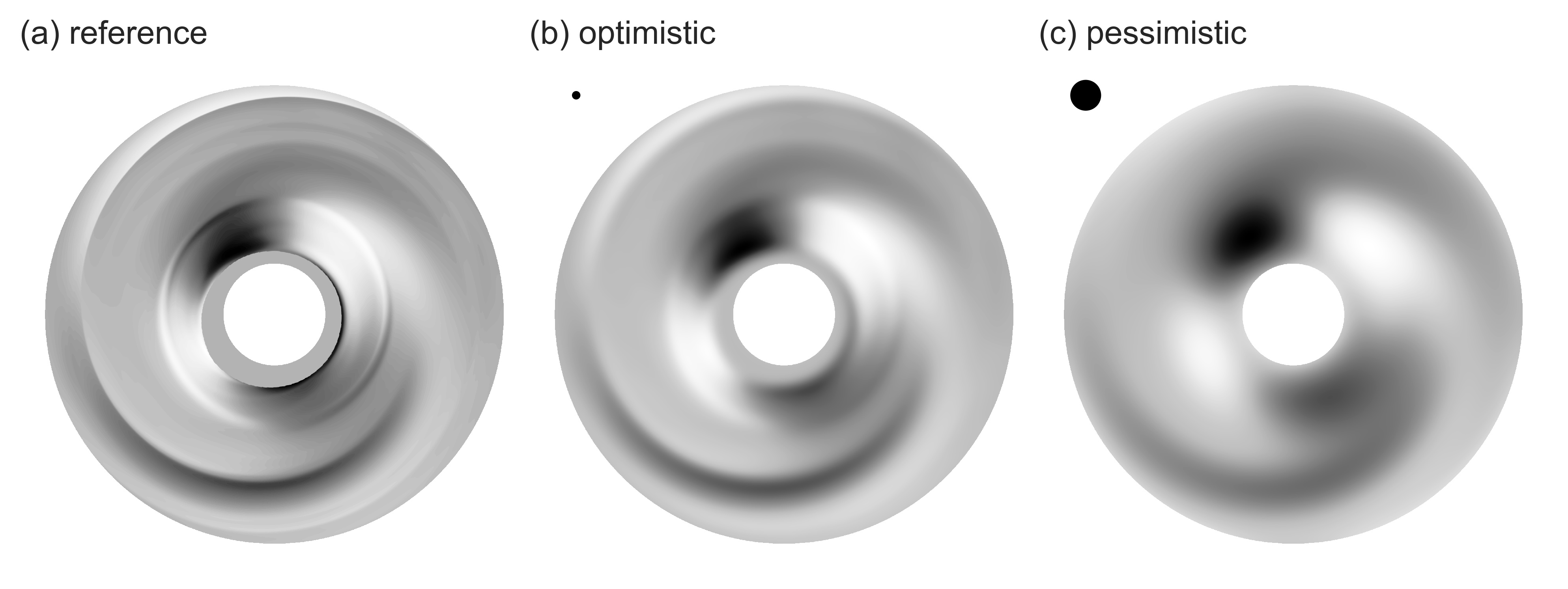}
    \caption{A qualitative comparison between a simulation-precision velocity
    map (left), and against artificially lowered spatial resolution, simulation
    with a gaussian kernel convolution. We apply kernels with angular size (in proportions
    of the target's) \SI{7}{\%} (center), and a 3 times larger one \SI{21}{\%} (left).
    With \hd's distance, the center panel corresponds to the recent high
    resolution obtained by \cite{Keppler2019a}. No noise is added. Projected
    velocities are shown in linear grey-scale, where $v'_\mathrm{LOS}<0$ is light and
    $v'_\mathrm{LOS}>0$ is dark grey. Secondary spiral patterns are lost at low resolution
    but the primary remains visible. Beamsize is shown as a black dot. The velocity map
    corresponds to the rightmost panel in \cref{fig:vproj_hd142_sprials}.}
    \label{fig:beamed_spiral}
\end{figure}

As shown in \cref{ssec:rwi_spirals}, the projected velocities seen in spiral arms are
comparable in amplitude to those attained by the vortex's core. However, angular
resolution might constitute an additional limitation to identify those secondary
structures. In \cref{fig:beamed_spiral} we mock a limited angular resolution via Gaussian
kernel convolution to the simulated velocity map, where the mean component of
azimuthal velocity $\langle v_\varphi\rangle$ is subtracted prior to projection. We
observe that the contrast sharpness of the main spiral pattern is altered but not
destroyed by limited spatial resolution alone. We note that the spiral arm appears
marginally broader in \cref{fig:proj_real} as compared with the numerical case
\cref{fig:proj_ideal}. The velocity flip pattern however remains visible and is unaltered
by the limited spatial resolution.

\subsection{The origin of the cavity in \hd}
The state of the art simulations for the thermal emission of \hd were performed in
Smooth-Particle Hydro (SPH) by \citet{Price2018}, and do not feature vortex formation.
This study was focused on explaining as many features as possible with the excitation
provided by the eccentric stellar companion.  However, it must be noted than SPH solvers
generate numerical viscosities $\sim 10^{-2}$ \citet{Arena2013}, much greater than typical
values used in RWI vortex studies\footnote{The model used in this paper is inviscid. We
give insight on our evaluation of numerical viscosity in \cref{app:numerical_viscosity}.}
\citep{Lyra2009, Hammer2017, Hammer2019, Ono2016}, so this possibility was inherently not
included in their study. In the present work, we stayed agnostic regarding how the initial
unstable density jump was formed. The stellar companion, while not included in our model,
provides a plausible cause to the cavity. However, gravitational perturber-induced Rossby
vortices have been studied in the context of circular orbital motion \citep{Li2005}. How
eccentricity and inclination in the companion's orbit affects vortices formation, within
an appropriately inviscid medium, remains to be studied.

\subsection{Limits of this approach}
An important limitation of the model is the lack of a vertical dimension. In a more
realistic context, plane velocities $(v_r, v_\varphi)$ are only detectable if the disk is
inclined, which will in turn affect measurements by line-of-sight integration. This effect
would however be mitigated by choosing optically thick molecular lines. Moreover,
\citep{Meheut2012a} showed that 3D vortices have a non negligible vertical velocity
component while they form (typically \SI{10}{\%} of the characteristic azimuthal
velocity signature). As the RWI growth time is typically shorter than the vortex lifetime
by one or two orders of magnitude, it seems reasonable to neglect vertical circulation.

It has been showed that the disk's contribution to the
gravitational potential, promotes disk eccentricity
\citep{Regaly2017}, which in turn amplifies the vortex's proper velocity.
Because this effect is neglected in our model, we expect the resulting velocities
to be
slightly under-estimated here.

\section{Conclusions}
\label{sec:ccl}
We showed that in cavity-hosting circumstellar disks, large eddies produce dynamical
signatures on the verge of detectability for current facilities.

As a vortex' dynamical imprint resides in the non-axisymmetric part of the velocity field,
it is crucial to the detection to be able to subtract the axisymmetric component from
observations.  In the case of a vortex formed at the inner edge of a cavity-hosting disk,
a Keplerian power-law is not a correct proxy for the mean azimuthal velocity. This is
because pressure gradients prone to vortex formation imply large deviations from keplerian
velocities. Nevertheless, as projected velocities of the observational major axis directly
map the azimuthal motion, a better mask can be obtained by averaging both sides of the
velocity profile on this axis. This approach proved to produce small errors when compared
with the actual azimuthal mean component of velocity $\vsym$. We also observed a
saturation in the amplitude of projected velocities as temperature is increased. This
result is to be taken with a grain of salt and may point to a limitation of the model we
used. Using this amplitude as an estimator for spectral resolution requirement, we
conclude that detection of a single large eddy is achievable under a
\SIrange{50}{150}{m.s^{-1}} resolution, while the current maximal resolution with ALMA is
$\sim$ \SI{30}{m.s^{-1}} We stress that those minimal requirements were obtained within
the particular case of the \hd target, with a relatively low inclination
(\SI{27}{\degree}). Minimal resolution would be amplified by a factor \SI{150}{\%} for a
more likely, mean inclination of \SI{45}{\degree}, ceteris paribus. This demanding
requirement may explain the current difficulty to elucidate the nature of known dust
clumps in cavity-hosting disks, yet is achievable with existing facilities. Vortex-free
mechanisms could also explain their formation, although observational constraints for fine
gas dynamics are needed in order to properly discriminate concurrent scenarios.

Full 3D modeling would naturally extend the present work, and allow the study of second
order effects in line-of-sight integration.
\begin{acknowledgements}
C.M.T. Robert's Ph.D. grant is part of ANR number ANR-16-CE31-0013 (Planet-Forming-Disks).
Computations were performed with \code{OCCIGEN} CINES (allocation \code{DARI
A0060402231}), and "Mésocentre SIGAMM", hosted by Observatoire de la Côte d'Azur. The
data presented in this work were proceeded and plotted with Python's data science ecosystem
\code{scipy}, \code{numpy} \citep{Walt2011}, \code{matplotlib} \citep{Hunter2007},
\code{pandas} \citep{McKinney2010}
as well as \code{astropy} \citep{Robitaille2013,Price-Whelan2018}, and the \code{yt}
framework \citep{Turk2011}.
C.M.T. is thankful to Aurélien Crida and Elena Lega for their early proofreading.
The authors wish to thank the referee, Zsolt Reg\`aly, for helping improving the quality of the paper.
\end{acknowledgements}

\appendix%
\section{Aspect ratios evaluations}
  
\subsection{Equivalence to locally isothermal}
\label{app:aspect_ratio}
Our model differs from the widely used locally isothermal prescription in that it is not
defined in terms of scale-height
\begin{equation}\label{eq:hflaring}
  H = h r (r/r_*)^\beta \:,
\end{equation}
where $h$ is the disk aspect ratio and $\beta$ is the flaring.  We can nonetheless draw an
equivalence with those parameters for the power law density distribution at the core of
\cref{eq:init_sigma}, such that $\Sigma(r) = \Sigma_0 (r/r_*)^{-1}$.  In the
locally isothermal prescription, the scale height is usually defined such that $H^2 =
c\sub{s}^2/\Omega_K^2$, so we can equate this with \cref{eq:hflaring} to get
\begin{equation}
  \begin{array}{ll}
    h^2 r^2 (r/r_*)^{2\beta} 
    &= \frac{\gamma p/\Sigma}{GM/r^3}\\
    &= \frac{\gamma S}{GM} r^3 \Sigma^{\gamma-1}\\
    &= \frac{\gamma S \Sigma_0^{\gamma-1}}{GM} r^3 (r/r\sub{j})^{1-\gamma}\:.
  \end{array}
\end{equation}
at which point we deduce an effective aspect ratio and disk flaring, in terms of the
actual simulation parameters
\begin{equation}\label{eq:h_and_beta}
  \left\{\begin{array}{ll}
  h^2 &= \frac{\gamma S \Sigma_0^{\gamma-1} r_*}{GM}
  \:,\\
  \beta &= 1 - \gamma/2 = 1/6\:.
  \end{array}\right.
\end{equation}

\begin{figure}[ht]
    \centering
    \includegraphics[width=\figwidth]{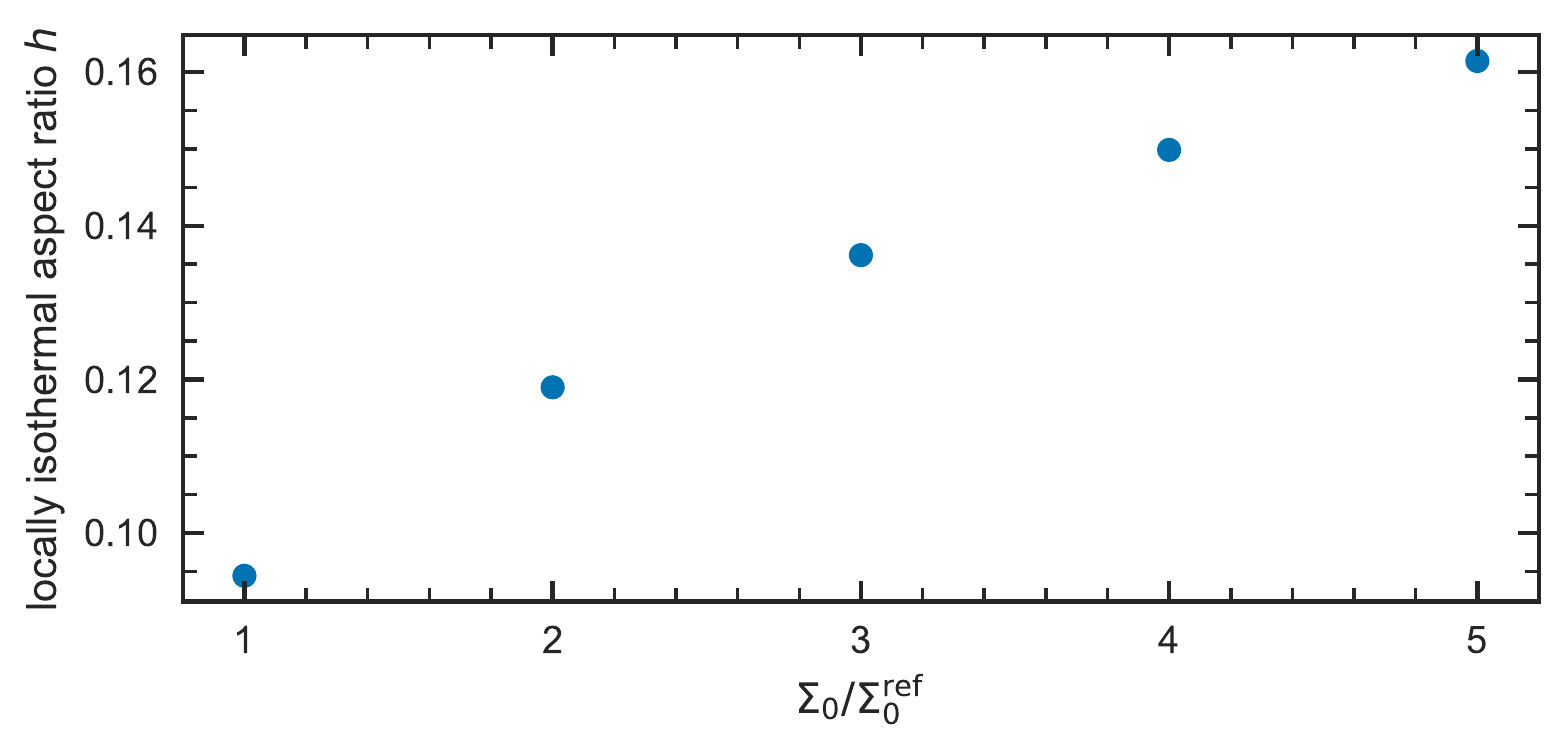}
    \caption{Correspondence between locally isothermal aspect ratio and "disk mass" in our
      simulations. The leftmost model is the reference.}
    \label{fig:lisoth_h}
\end{figure}

We note that our fixed resolution corresponds to $\Delta r/H(r\sub{j})
\simeq 0.04$ for the reference model.  \Cref{fig:lisoth_h}
shows the resulting variation in $h$ as we scale up $\Sigma_0$, following
\cref{eq:h_and_beta}.

\subsection{Spiral fitting}
\begin{figure}
  \centering
  \includegraphics[width=\figwidth]{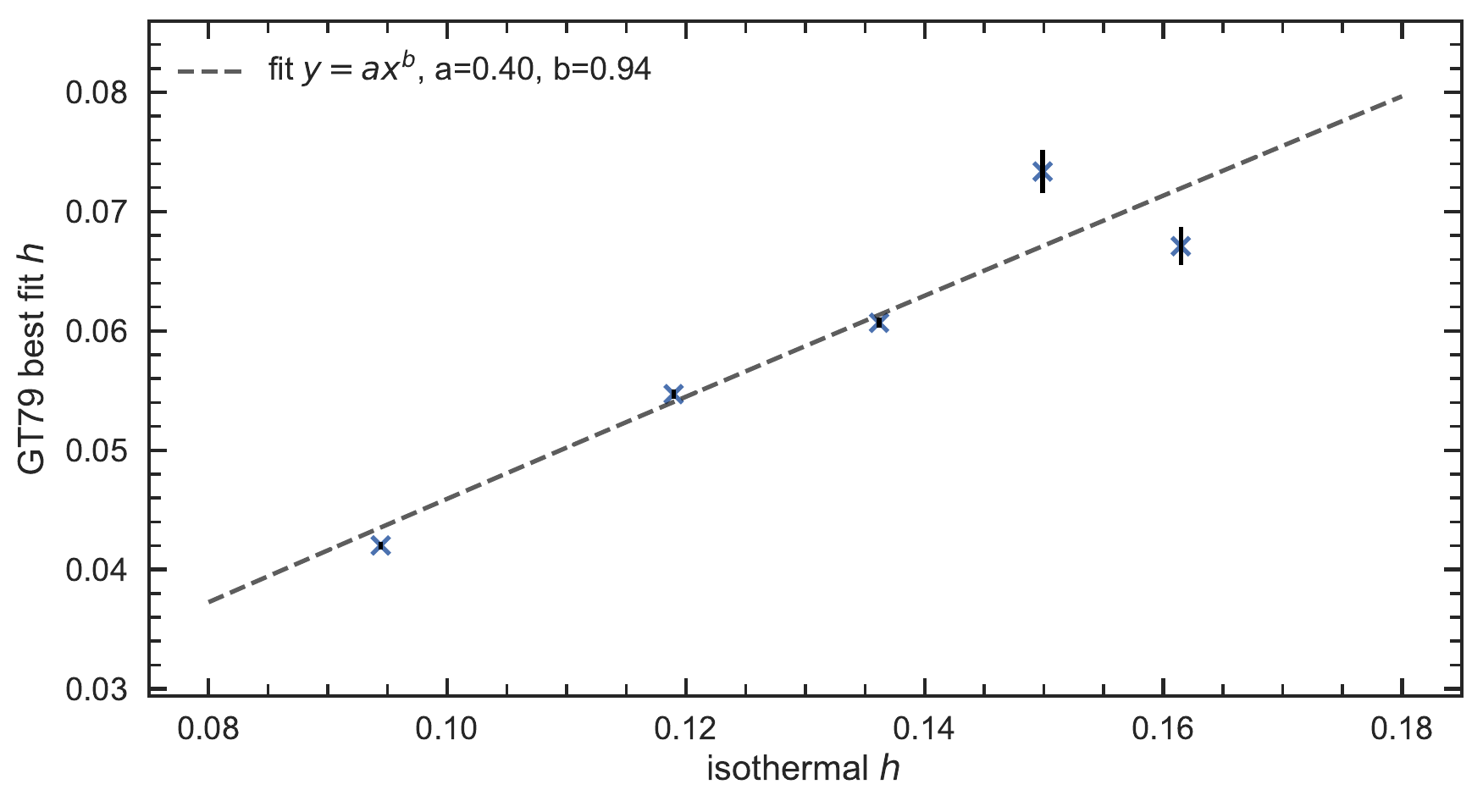}
  \caption{Aspect ratios as defined in \cref{app:aspect_ratio} VS empirical
    values obtained from fitting \Cref{eq:linear_regime_spiral}. The later is roughly
    $40\%$ of the former.}
  \label{fig:GT79_vs_lisoth_h}
\end{figure}

In \cref{fig:vproj_hd142_sprials}, we fitted the linear-regime spiral wave shape
\citep{Goldreich1979,Rafikov2002,Muto2012} given by
\begin{equation}
\begin{split}\label{eq:linear_regime_spiral}
  &\varphi(r) = \varphi_o - \frac{\mathrm{sgn}(r-r_o)}{H_o} \times
  \\&\left((r/r_o)^{1+\beta}
      \left[\frac{1}{1+\beta}-\frac{1}{1-\alpha+\beta}(r/r_o)^{-\alpha}\right]
  - \left[\frac{1}{1+\beta}-\frac{1}{1-\alpha+\beta}\right]\right)
    \:,
\end{split}
\end{equation}
where $\alpha, \beta$ are power-law exponents respectively defined as $\Omega
\propto r^{-\alpha}$ and $c_\mathrm{s} \propto r^{-\beta}$. $(r_o, \varphi_o)$
are the spiral origin's coordinates, while $H_o$ is a scale-height at this
position. The fit was performed with $H_o$ as a free parameter, so we the corresponding
aspect ratio, differs from the locally isothermal equivalent $h$ value used throughout the
paper and described in the previous section. \Cref{fig:GT79_vs_lisoth_h} shows values
against each other.

\section{Numerical viscosity evaluation}
\label{app:numerical_viscosity}
\begin{figure}
  \centering
  \includegraphics[width=\figwidth]{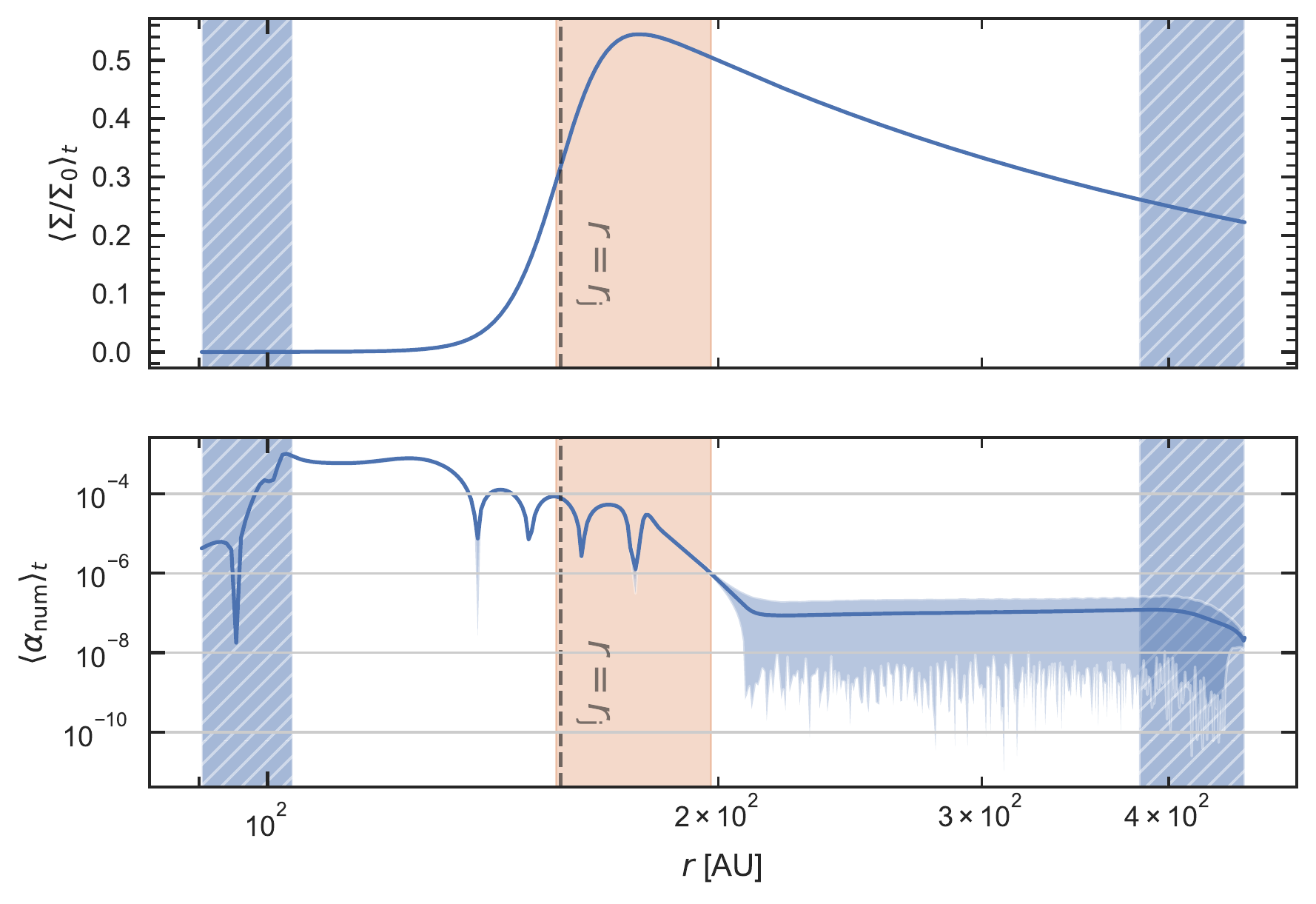}
  \caption{Density (top) and numerical viscosity equivalent $\alpha$ value
  (bottom) time-averaged over 10 orbital periods ($t/t_*\in[90, 100]$) with a sampling
  rate of $0.1$ orbital periods. The solid blue shadows indicate the variation interval
  over the sample timeseries, showing that the profile is very stable in the region of
  interest. Hatched regions highlight the wave killing zones, while the orange region
  loosely indicates the vortex forming region, spanning one scale height away from the
  local density maximum.}
  \label{fig:alpha_visc_eval}
\end{figure}

In order to estimate numerical viscosity $\nu\sub{num}(r)$, performed a 1D in a
1D run, with identical parameterization as our reference 2D run (run1).
The analytical initial conditions constitute a stable equilibrium since RWI can not grow
in 1D. Since our boundary conditions do not impose mass flux, any radial mass transport
$\dot{M}$ through the simulation domain is caused by numerical viscosity such that
$\nu\sub{num} \Sigma = |\dot{M}|/3\pi$. In terms of \citet{Shakura1973}'s alpha viscosity
model $\nu\sub{num} =
\alpha\sub{num} H c\sub{s}$, so finally
\begin{equation}
    \alpha\sub{num} = \frac{2}{3} \left| \frac{v_r}{hc\sub{s}} \right|\:.
\end{equation}
The obtained profile, time-averaged, is plotted in \cref{fig:alpha_visc_eval}.
Highest numerical viscosities ($\sim 2\times 10^{-3}$) are reached in the cavity, while it
stays bounded $<10^{-4}$ in the vortex-forming region, roughly represented in orange.


\begin{thebibliography}{83}
\expandafter\ifx\csname natexlab\endcsname\relax\def\natexlab#1{#1}\fi

\bibitem[{Adams \& Watkins(1995)}]{Adams1995}
Adams, F.~C. \& Watkins, R. 1995, ApJ, 451, 314

\bibitem[{Andrews {et~al.}(2018)Andrews, Terrell, Tripathi, Ansdell, Williams,
  \& Wilner}]{Andrews2018}
Andrews, S.~M., Terrell, M., Tripathi, A., {et~al.} 2018, ApJ, 865, 157

\bibitem[{Arena \& Gonzalez(2013)}]{Arena2013}
Arena, S.~E. \& Gonzalez, J.-F. 2013, Mon Not R Astron Soc, 433, 98

\bibitem[{Ataiee {et~al.}(2014)Ataiee, Dullemond, Kley, Reg{\'a}ly, \&
  Meheut}]{Ataiee2014}
Ataiee, S., Dullemond, C.~P., Kley, W., Reg{\'a}ly, Z., \& Meheut, H. 2014,
  A\&A, 572, A61

\bibitem[{Ataiee {et~al.}(2013)Ataiee, Pinilla, Zsom, Dullemond, Dominik, \&
  Ghanbari}]{Ataiee2013}
Ataiee, S., Pinilla, P., Zsom, A., {et~al.} 2013, A\&A, 553, L3

\bibitem[{Avenhaus {et~al.}(2014)Avenhaus, Quanz, Schmid, Meyer, Garufi, Wolf,
  \& Dominik}]{Avenhaus2014}
Avenhaus, H., Quanz, S.~P., Schmid, H.~M., {et~al.} 2014, ApJ, 781, 87

\bibitem[{Barge \& Sommeria(1995)}]{Barge1995}
Barge, P. \& Sommeria, J. 1995, A\&A, 295, L1

\bibitem[{Baruteau {et~al.}(2019)Baruteau, Barraza, P{\'e}rez, Casassus, Dong,
  Lyra, Marino, Christiaens, Zhu, Carmona, Debras, \& Alarcon}]{Baruteau2019}
Baruteau, C., Barraza, M., P{\'e}rez, S., {et~al.} 2019, MNRAS, 486, 304

\bibitem[{Benisty {et~al.}(2018)Benisty, Juhasz, Facchini, Pinilla, {de Boer},
  P{\'e}rez, Keppler, {Muro-Arena}, Villenave, Andrews, Dominik, Dullemond,
  Gallenne, Garufi, Ginski, \& Isella}]{Benisty2018}
Benisty, M., Juhasz, A., Facchini, S., {et~al.} 2018, A\&A, 619, 171

\bibitem[{Biller {et~al.}(2012)Biller, Lacour, Juhasz, Benisty, Chauvin,
  Olofsson, Pott, Uller, {Sicilia-Aguilar}, Bonnefoy, Tuthill, Thebault,
  Henning, \& Crida}]{Biller2012}
Biller, B., Lacour, S., Juhasz, A., {et~al.} 2012, ApJ Lett., 753, L38

\bibitem[{Birnstiel {et~al.}(2013)Birnstiel, Dullemond, \&
  Pinilla}]{Birnstiel2013}
Birnstiel, T., Dullemond, C.~P., \& Pinilla, P. 2013, A\&A, 550, L8

\bibitem[{Boehler {et~al.}(2017)Boehler, Weaver, Isella, Ricci, Grady,
  Carpenter, \& Perez}]{Boehler2017}
Boehler, Y., Weaver, E., Isella, A., {et~al.} 2017, ApJ, 840, 60

\bibitem[{Bracco {et~al.}(1999)Bracco, Chavanis, Provenzale, \&
  Spiegel}]{Bracco1999}
Bracco, A., Chavanis, P.~H., Provenzale, A., \& Spiegel, E.~A. 1999, Physics of
  Fluids, 11, 2280

\bibitem[{Casassus {et~al.}(2019)Casassus, Marino, Lyra, Baruteau, Vidal,
  Wootten, P{\'e}rez, Alarcon, Barraza, C{\'a}rcamo, Dong, Sierra, Zhu, Ricci,
  Christiaens, \& Cieza}]{Casassus2019a}
Casassus, S., Marino, S., Lyra, W., {et~al.} 2019, MNRAS, 483, 3278

\bibitem[{Casassus {et~al.}(2015{\natexlab{a}})Casassus, Marino, P{\'e}rez,
  Roman, Dunhill, Armitage, Cuadra, Wootten, {van der Plas}, Cieza, Moral,
  Christiaens, \& Montesinos}]{Casassus2015a}
Casassus, S., Marino, S., P{\'e}rez, S., {et~al.} 2015{\natexlab{a}}, ApJ, 811,
  92

\bibitem[{Casassus \& P{\'e}rez(2019)}]{Casassus2019}
Casassus, S. \& P{\'e}rez, S. 2019, The Astrophysical Journal Letters, 883, L41

\bibitem[{Casassus {et~al.}(2012)Casassus, Perez~M., Jord{\'a}n, M{\'e}nard,
  Cuadra, Schreiber, Hales, \& Ercolano}]{Casassus2012}
Casassus, S., Perez~M., S., Jord{\'a}n, A., {et~al.} 2012, ApJ Lett., 754, L31

\bibitem[{Casassus {et~al.}(2015{\natexlab{b}})Casassus, Wright, Marino,
  Maddison, Wootten, Rom{\'a}n, Perez, Pinilla, Wyatt, Moral, M{\'e}nard,
  Christiaens, Cieza, \& {van der Plas}}]{Casassus2015b}
Casassus, S., Wright, C., Marino, S., {et~al.} 2015{\natexlab{b}}, ApJ, 812,
  126

\bibitem[{Cazzoletti {et~al.}(2018)Cazzoletti, {van Dishoeck}, Pinilla,
  Tazzari, Facchini, {van der Marel}, Benisty, Garufi, \&
  P{\'e}rez}]{Cazzoletti2018}
Cazzoletti, P., {van Dishoeck}, E.~F., Pinilla, P., {et~al.} 2018, A\&A, 619,
  A161

\bibitem[{Chiang \& Youdin(2010)}]{Chiang2010a}
Chiang, E. \& Youdin, A. 2010, Annu. Rev. Earth Planet. Sci., 38, 493

\bibitem[{Christiaens {et~al.}(2014)Christiaens, Casassus, Perez, {van der
  Plas}, \& M{\'e}nard}]{Christiaens2014}
Christiaens, V., Casassus, S., Perez, S., {van der Plas}, G., \& M{\'e}nard, F.
  2014, ApJL, 785, L12

\bibitem[{{de Val-Borro} {et~al.}(2007){de Val-Borro}, Artymowicz, D'Angelo, \&
  Peplinski}]{deValBorro2007}
{de Val-Borro}, M., Artymowicz, P., D'Angelo, G., \& Peplinski, A. 2007, A\&A,
  471, 1043

\bibitem[{{de Val-Borro} {et~al.}(2006){de Val-Borro}, Edgar, Artymowicz,
  Ciecielag, Cresswell, D'Angelo, {Delgado-Donate}, Dirksen, Fromang,
  Gawryszczak, Klahr, Kley, Lyra, Masset, Mellema, Nelson, Paardekooper,
  Peplinski, Pierens, Plewa, Rice, Schaefer, \& Speith}]{deValBorro2006}
{de Val-Borro}, M., Edgar, R.~G., Artymowicz, P., {et~al.} 2006, MNRAS, 370,
  529

\bibitem[{Dong {et~al.}(2018)Dong, Liu, Eisner, Andrews, Fung, Zhu, Chiang,
  Hashimoto, Liu, Casassus, Esposito, Hasegawa, Muto, Pavlyuchenkov, Wilner,
  Akiyama, Tamura, \& Wisniewski}]{Dong2018}
Dong, R., Liu, S.-Y., Eisner, J., {et~al.} 2018, ApJ, 860, 124

\bibitem[{Fu {et~al.}(2014)Fu, Li, Lubow, \& Li}]{Fu2014a}
Fu, W., Li, H., Lubow, S., \& Li, S. 2014, ApJ Lett., 788, L41

\bibitem[{Fukagawa {et~al.}(2006)Fukagawa, Tamura, Itoh, Kudo, Imaeda, Oasa,
  Hayashi, \& Hayashi}]{Fukagawa2006}
Fukagawa, M., Tamura, M., Itoh, Y., {et~al.} 2006, ApJ, 636, L153

\bibitem[{Fukagawa {et~al.}(2013)Fukagawa, Tsukagoshi, Momose, Saigo, Ohashi,
  Kitamura, Inutsuka, Muto, Nomura, Takeuchi, Kobayashi, Hanawa, Akiyama,
  Honda, Fujiwara, Kataoka, Takahashi, \& Shibai}]{Fukagawa2013}
Fukagawa, M., Tsukagoshi, T., Momose, M., {et~al.} 2013, Publ Astron Soc Jpn
  Nihon Tenmon Gakkai, 65

\bibitem[{{Gaia Collaboration} {et~al.}(2016){Gaia Collaboration}, Brown,
  Vallenari, Prusti, {de Bruijne}, Mignard, \& Drimmel}]{Gaia2016}
{Gaia Collaboration}, Brown, A. G.~A., Vallenari, A., {et~al.} 2016, A\&A, 595

\bibitem[{Goldreich \& Tremaine(1979)}]{Goldreich1979}
Goldreich, P. \& Tremaine, S. 1979, The Astrophysical Journal, 233, 857

\bibitem[{Hammer {et~al.}(2017)Hammer, Kratter, \& Lin}]{Hammer2017}
Hammer, M., Kratter, K.~M., \& Lin, M.-K. 2017, MNRAS, 466, 3533

\bibitem[{Hammer {et~al.}(2019)Hammer, Pinilla, Kratter, \& Lin}]{Hammer2019}
Hammer, M., Pinilla, P., Kratter, K.~M., \& Lin, M.-K. 2019, MNRAS, 482, 3609

\bibitem[{Harten {et~al.}(1983)Harten, Lax, \& van. Leer}]{Harten1983}
Harten, A., Lax, P.~D., \& van. Leer, B. 1983, SIAM Rev., 25, 35

\bibitem[{Huang {et~al.}(2019)Huang, Dong, Li, Li, \& Ji}]{Huang2019}
Huang, P., Dong, R., Li, H., Li, S., \& Ji, J. 2019, ApJ, 883, L39

\bibitem[{Hunter(2007)}]{Hunter2007}
Hunter, J.~D. 2007, Comput. Sci. Eng., 9, 99

\bibitem[{Isella {et~al.}(2018)Isella, Huang, Andrews, Dullemond, Birnstiel,
  Zhang, Zhu, Guzm{\'a}n, P{\'e}rez, Bai, Benisty, Carpenter, Ricci, \&
  Wilner}]{Isella2018}
Isella, A., Huang, J., Andrews, S.~M., {et~al.} 2018, ApJL, 869, L49

\bibitem[{Kataoka {et~al.}(2016)Kataoka, Tsukagoshi, Momose, Nagai, Muto,
  Dullemond, Pohl, Fukagawa, Shibai, Hanawa, \& Murakawa}]{Kataoka2016}
Kataoka, A., Tsukagoshi, T., Momose, M., {et~al.} 2016, ApJ Lett., 831

\bibitem[{Keppler {et~al.}(2019)Keppler, Teague, Bae, Benisty, Henning, van
  Boekel, Chapillon, Pinilla, Williams, Bertrang, Facchini, Flock, Ginski,
  Juhasz, Klahr, Liu, M{\"u}ller, P{\'e}rez, Pohl, Rosotti, Samland, \&
  Semenov}]{Keppler2019a}
Keppler, M., Teague, R., Bae, J., {et~al.} 2019, A\&A, 625, A118

\bibitem[{Koren(1993)}]{Koren1993}
Koren, B. 1993, Rep.-Dep. Numer. Math., 1

\bibitem[{Lai \& Tsang(2009)}]{Lai2009}
Lai, D. \& Tsang, D. 2009, MNRAS, 393, 979

\bibitem[{Li {et~al.}(2001)Li, Colgate, Wendroff, \& Liska}]{Li2001}
Li, H., Colgate, S.~A., Wendroff, B., \& Liska, R. 2001, ApJ, 551, 874

\bibitem[{Li {et~al.}(2000)Li, Finn, Lovelace, \& Colgate}]{Li2000}
Li, H., Finn, J.~M., Lovelace, R. V.~E., \& Colgate, S.~A. 2000, ApJ, 533, 1023

\bibitem[{Li {et~al.}(2005)Li, Li, Koller, Wendroff, Liska, Orban, Liang, \&
  Lin}]{Li2005}
Li, H., Li, S., Koller, J., {et~al.} 2005, ApJ, 624, 1003

\bibitem[{Lin(2012)}]{Lin2012}
Lin, M. 2012, AGU Fall Meeting Abstracts, 21, P21B

\bibitem[{Lovelace {et~al.}(1999)Lovelace, Li, Colgate, \&
  Nelson}]{Lovelace1999}
Lovelace, R. V.~E., Li, H., Colgate, S.~A., \& Nelson, A.~F. 1999, ApJ, 513,
  805

\bibitem[{Lyra {et~al.}(2009)Lyra, Johansen, Klahr, \& {Piskunov}}]{Lyra2009}
Lyra, W., Johansen, A., Klahr, H., \& {Piskunov}. 2009, A\&A, 493, 1125

\bibitem[{McKinney(2010)}]{McKinney2010}
McKinney, W. 2010, in Proceedings of the 9th {{Python}} in {{Science
  Conference}}, 51--56

\bibitem[{McNally {et~al.}(2018)McNally, Nelson, \& Paardekooper}]{Mcnally2018}
McNally, C.~P., Nelson, R.~P., \& Paardekooper, S.-J. 2018, MNRAS, 477, 4596

\bibitem[{M{\'e}heut {et~al.}(2012{\natexlab{a}})M{\'e}heut, Keppens, Casse, \&
  Benz}]{Meheut2012b}
M{\'e}heut, H., Keppens, R., Casse, F., \& Benz, W. 2012{\natexlab{a}}, A\&A,
  542, A9

\bibitem[{M{\'e}heut {et~al.}(2012{\natexlab{b}})M{\'e}heut, Yu, \&
  Lai}]{Meheut2012a}
M{\'e}heut, H., Yu, C., \& Lai, D. 2012{\natexlab{b}}, MNRAS, 422, 2399

\bibitem[{Montesinos {et~al.}(2016)Montesinos, Perez, Casassus, Marino, Cuadra,
  \& Christiaens}]{Montesinos2016}
Montesinos, M., Perez, S., Casassus, S., {et~al.} 2016, Astrophys. J. Lett.,
  823, L8

\bibitem[{Muto {et~al.}(2012)Muto, Grady, Hashimoto, Fukagawa, Hornbeck, Sitko,
  Russell, Werren, Cur{\'e}, Currie, Ohashi, Okamoto, Momose, Honda, Inutsuka,
  Takeuchi, Dong, Abe, Brandner, Brandt, Carson, Egner, Feldt, Fukue, Goto,
  Guyon, Hayano, Hayashi, Hayashi, Henning, Hodapp, Ishii, Iye, Janson,
  Kandori, Knapp, Kudo, Kusakabe, Kuzuhara, Matsuo, Mayama, McElwain, Miyama,
  Morino, {Moro-Martin}, Nishimura, Pyo, Serabyn, Suto, Suzuki, Takami, Takato,
  Terada, Thalmann, Tomono, Turner, Watanabe, Wisniewski, Yamada, Takami,
  Usuda, \& Tamura}]{Muto2012}
Muto, T., Grady, C.~A., Hashimoto, J., {et~al.} 2012, ApJL, 748, L22

\bibitem[{Ono {et~al.}(2016)Ono, Muto, Takeuchi, \& Nomura}]{Ono2016}
Ono, T., Muto, T., Takeuchi, T., \& Nomura, H. 2016, ApJ, 823, 84

\bibitem[{P{\'e}rez {et~al.}(2020)P{\'e}rez, Casassus, Hales, Marino, Cheetham,
  Zurlo, Cieza, Dong, Alarc{\'o}n, {Ben{\'i}tez-Llambay}, Fomalont, \&
  Avenhaus}]{Perez2020}
P{\'e}rez, S., Casassus, S., Hales, A., {et~al.} 2020, ApJL, 889, L24

\bibitem[{Perez {et~al.}(2015)Perez, Casassus, M{\'e}nard, Roman, Van Der~Plas,
  Cieza, Pinte, Christiaens, \& Hales}]{Perez2015}
Perez, S., Casassus, S., M{\'e}nard, F., {et~al.} 2015, ApJ, 798, 85

\bibitem[{Pineda {et~al.}(2019)Pineda, Szul{\'a}gyi, Quanz, Van~Dishoeck,
  Garufi, Meru, Mulders, Testi, Meyer, \& Reggiani}]{Pineda2019}
Pineda, J.~E., Szul{\'a}gyi, J., Quanz, S.~P., {et~al.} 2019, ApJ, 871, 48

\bibitem[{Pinte {et~al.}(2018)Pinte, Price, M{\'e}nard, Duchene, Dent, Hill,
  {De Gregorio-Monsalvo}, Hales, \& Mentiplay}]{Pinte2018}
Pinte, C., Price, D.~J., M{\'e}nard, F., {et~al.} 2018, ApJ Lett., 860, L13

\bibitem[{Pinte {et~al.}(2019)Pinte, van~der Plas, M{\'e}nard, Price,
  Christiaens, Hill, Mentiplay, Ginski, Choquet, Boehler, Duch{\^e}ne, Perez,
  \& Casassus}]{Pinte2019a}
Pinte, C., van~der Plas, G., M{\'e}nard, F., {et~al.} 2019, Nat Astron, 1

\bibitem[{Porth {et~al.}(2014)Porth, Xia, Hendrix, Moschou, \&
  Keppens}]{Porth2014}
Porth, O., Xia, C., Hendrix, T., Moschou, S.~P., \& Keppens, R. 2014, ApJ Suppl
  Ser, 214, 4

\bibitem[{Price {et~al.}(2018)Price, Cuello, Pinte, Mentiplay, Casassus,
  Christiaens, Kennedy, Cuadra, Sebastian~Perez, Marino, Armitage, Zurlo,
  Juhasz, Ragusa, Laibe, \& Lodato}]{Price2018}
Price, D.~J., Cuello, N., Pinte, C., {et~al.} 2018, MNRAS, 477, 1270

\bibitem[{{Price-Whelan} {et~al.}(2018){Price-Whelan}, Sip{\textbackslash}Hocz,
  G{\"u}nther, Lim, Crawford, Conseil, Shupe, Craig, Dencheva, Ginsburg,
  VanderPlas, Bradley, {P{\'e}rez-Su{\'a}rez}, de~{Val-Borro}, Aldcroft, Cruz,
  Robitaille, Tollerud, Ardelean, Babej, Bach, Bachetti, Bakanov, Bamford,
  Barentsen, Barmby, Baumbach, Berry, Biscani, Boquien, Bostroem, Bouma,
  Brammer, Bray, Breytenbach, Buddelmeijer, Burke, Calderone, Rodr{\'i}guez,
  Cara, Cardoso, Cheedella, Copin, Corrales, Crichton, D'Avella, Deil, Depagne,
  Dietrich, Donath, Droettboom, Earl, Erben, Fabbro, Ferreira, Finethy, Fox,
  Garrison, Gibbons, Goldstein, Gommers, Greco, Greenfield, Groener, Grollier,
  Hagen, Hirst, Homeier, Horton, Hosseinzadeh, Hu, Hunkeler, Ivezi{\'c}, Jain,
  Jenness, Kanarek, Kendrew, Kern, Kerzendorf, Khvalko, King, Kirkby, Kulkarni,
  Kumar, Lee, Lenz, Littlefair, Ma, Macleod, Mastropietro, McCully, Montagnac,
  Morris, Mueller, Mumford, Muna, Murphy, Nelson, Nguyen, Ninan, N{\"o}the,
  Ogaz, Oh, Parejko, Parley, Pascual, Patil, Patil, Plunkett, Prochaska,
  Rastogi, Janga, Sabater, Sakurikar, Seifert, Sherbert, {Sherwood-Taylor},
  Shih, Sick, Silbiger, Singanamalla, Singer, Sladen, Sooley, Sornarajah,
  Streicher, Teuben, Thomas, Tremblay, Turner, Terr{\'o}n, van Kerkwijk, de~la
  Vega, Watkins, Weaver, Whitmore, Woillez, Zabalza, \&
  {and}}]{Price-Whelan2018}
{Price-Whelan}, a. A.~M., Sip{\textbackslash}Hocz, B.~M., G{\"u}nther, H.~M.,
  {et~al.} 2018, AJ, 156, 123

\bibitem[{Rafikov(2002)}]{Rafikov2002}
Rafikov, R.~R. 2002, ApJ, 572, 566

\bibitem[{Rameau {et~al.}(2012)Rameau, Chauvin, Lagrange, Th{\'e}bault, Milli,
  Girard, \& Bonnefoy}]{Rameau2012}
Rameau, J., Chauvin, G., Lagrange, A.-M., {et~al.} 2012, A\&A, 546, A24

\bibitem[{Rayleigh(1879)}]{Rayleigh1879}
Rayleigh, J.~W. 1879, Proc. Lond. Math. Soc., 1, 57

\bibitem[{Reg{\'a}ly {et~al.}(2012)Reg{\'a}ly, Juhasz, S{\'a}ndor, \&
  Dullemond}]{Regaly2012}
Reg{\'a}ly, Z., Juhasz, A., S{\'a}ndor, Z., \& Dullemond, C.~P. 2012, MNRAS,
  419, 1701

\bibitem[{Reg{\'a}ly {et~al.}(2013)Reg{\'a}ly, S{\'a}ndor, Csom{\'o}s, \&
  Ataiee}]{Regaly2013}
Reg{\'a}ly, Z., S{\'a}ndor, Z., Csom{\'o}s, P., \& Ataiee, S. 2013, MNRAS, 433,
  2626

\bibitem[{Reg{\'a}ly \& Vorobyov(2017{\natexlab{a}})}]{Regaly2017}
Reg{\'a}ly, Z. \& Vorobyov, E. 2017{\natexlab{a}}, A\&A, 601, A24

\bibitem[{Reg{\'a}ly \& Vorobyov(2017{\natexlab{b}})}]{Regaly2017a}
Reg{\'a}ly, Z. \& Vorobyov, E. 2017{\natexlab{b}}, MNRAS, 471, 2204

\bibitem[{Richard {et~al.}(2013)Richard, Barge, \& Le~Diz{\`e}s}]{Richard2013}
Richard, S., Barge, P., \& Le~Diz{\`e}s, S. 2013, A\&A, 559, A30

\bibitem[{Robitaille {et~al.}(2013)Robitaille, Tollerud, Greenfield,
  Droettboom, Bray, Aldcroft, Davis, Ginsburg, {Price-Whelan}, Kerzendorf,
  Conley, Crighton, Barbary, Muna, Ferguson, Grollier, Parikh, Nair,
  G{\"u}nther, Deil, Woillez, Conseil, Kramer, Turner, Singer, Fox, Weaver,
  Zabalza, Edwards, Bostroem, Burke, Casey, Crawford, Dencheva, Ely, Jenness,
  Labrie, Lim, Pierfederici, Pontzen, Ptak, Refsdal, Servillat, \&
  Streicher}]{Robitaille2013}
Robitaille, T.~P., Tollerud, E.~J., Greenfield, P., {et~al.} 2013, A\&A, 558,
  A33

\bibitem[{Shakura \& Sunyaev(1973)}]{Shakura1973}
Shakura, N.~I. \& Sunyaev, R.~A. 1973, A\&A, 500, 33

\bibitem[{Tanga {et~al.}(1996)Tanga, Babiano, \& Dubrulle}]{Tanga1996}
Tanga, P., Babiano, A., \& Dubrulle, B. 1996, Icarus, 121, 158

\bibitem[{Teague {et~al.}(2018{\natexlab{a}})Teague, Bae, Bergin, Birnstiel, \&
  {Foreman-Mackey}}]{Teague2018a}
Teague, R., Bae, J., Bergin, E.~A., Birnstiel, T., \& {Foreman-Mackey}, D.
  2018{\natexlab{a}}, ApJL, 860, L12

\bibitem[{Teague {et~al.}(2018{\natexlab{b}})Teague, Bae, Birnstiel, \&
  Bergin}]{Teague2018b}
Teague, R., Bae, J., Birnstiel, T., \& Bergin, E.~A. 2018{\natexlab{b}}, ApJ,
  868, 113

\bibitem[{Teague \& {Foreman-Mackey}(2018)}]{Teague2018}
Teague, R. \& {Foreman-Mackey}, D. 2018, Res. Notes Am. Astron. Soc., 2, 173

\bibitem[{Teague {et~al.}(2016)Teague, Guilloteau, Semenov, Henning, Dutrey,
  Pi{\'e}tu, Birnstiel, Chapillon, Hollenbach, \& Gorti}]{Teague2016}
Teague, R., Guilloteau, S., Semenov, D., {et~al.} 2016, A\&A, 592, A49

\bibitem[{Turk {et~al.}(2011)Turk, Smith, Oishi, Skory, Skillman, Abel, \&
  Norman}]{Turk2011}
Turk, M.~J., Smith, B.~D., Oishi, J.~S., {et~al.} 2011, ApJ Suppl. Ser, 192, 9

\bibitem[{{van der Marel} {et~al.}(2016){van der Marel}, {van Dishoeck},
  Bruderer, Andrews, Pontoppidan, Herczeg, {van Kempen}, \&
  Miotello}]{VanDerMarel2016a}
{van der Marel}, N., {van Dishoeck}, E.~F., Bruderer, S., {et~al.} 2016, A\&A,
  585, A58

\bibitem[{van~der Walt {et~al.}(2011)van~der Walt, Colbert, \&
  Varoquaux}]{Walt2011}
van~der Walt, S., Colbert, S.~C., \& Varoquaux, G. 2011, Comput. Sci. Eng., 13,
  22

\bibitem[{Verhoeff {et~al.}(2011)Verhoeff, Min, Pantin, Waters, Tielens, Honda,
  Fujiwara, Bouwman, {van Boekel}, Dougherty, De~Koter, Dominik, \&
  Mulders}]{Verhoeff2011}
Verhoeff, A.~P., Min, M., Pantin, E., {et~al.} 2011, A\&A, 528, A91

\bibitem[{Xia {et~al.}(2018)Xia, Teunissen, Mellah, Chan{\'e}, \&
  Keppens}]{Xia2018}
Xia, C., Teunissen, J., Mellah, I.~E., Chan{\'e}, E., \& Keppens, R. 2018, ApJ
  Suppl Ser, 234, 30

\bibitem[{Yen {et~al.}(2016)Yen, Liu, Gu, Hirano, Lee, Puspitaningrum, \&
  Takakuwa}]{Yen2016}
Yen, H.-W., Liu, H.~B., Gu, P.-G., {et~al.} 2016, ApJ Lett., 820, L25

\bibitem[{Zhu \& Baruteau(2016)}]{Zhu2016}
Zhu, Z. \& Baruteau, C. 2016, MNRAS, 458, 3918

\bibitem[{Zhu \& Stone(2014)}]{Zhu2014}
Zhu, Z. \& Stone, J.~M. 2014, ApJ, 795, 53

\end{thebibliography}
\end{document}